\documentclass[pop,fleqn]{w-art}
\usepackage{times}
\usepackage{amsmath}
\usepackage{w-thm}
\usepackage{bm}
\usepackage[dvips]{graphicx}
\begin{document}
\DOIsuffix{theDOIsuffix}
\Volume{51}
\Issue{1}
\Month{01}
\Year{2003}
\pagespan{3}{}



\title[Full Counting Statistics of Interacting Electrons]{Full Counting Statistics of Interacting Electrons.}


\author[D.~A. Bagrets]{D.~A. Bagrets\inst{1,2}}
\address[\inst{1}]{Institut f\"{u}r Theoretische Festk\"{o}perphysik
and DFG-Center for Functional Nanostructures (CFN), \\
Universit\"{a}t Karlsruhe, 76128 Karlsruhe, Germany}
\address[\inst{2}]{Forschungszentrum Karlsruhe, Institut f\"ur Nanotechnologie,
76021 Karlsruhe, Germany}
\author[Y. Utsumi]{Y. Utsumi\inst{1,3}}
\address[\inst{3}]{Condensed Matter Theory Laboratory, 
The Institute of Physical and Chemical Research (RIKEN),\\ 
Wako, Saitama 351-0198, Japan}
\author[D.~S.~Golubev]{D.~S.~Golubev\inst{1,2,4}}
\address[\inst{4}]{I.E. Tamm Department of Theoretical Physics, 
P.N. Lebedev Physics Institute, 119991 Moscow, Russia}
\author[G.~Sch\"on]{Gerd~Sch\"on\inst{1,2}}

\begin{abstract}
In order to fully characterize the noise associated with electron transport, 
with its severe consequences for solid-state quantum information systems,
the theory of full counting statistics has been developed.
It accounts for correlation effects associated with the statistics and 
effects of entanglement, but it remains a non-trivial task to account for 
interaction effects. 
In this article we present two examples: we describe electron transport 
through quantum dots with strong charging effects beyond perturbation 
theory in the tunneling, and we analyze current fluctuations in a diffusive 
interacting conductor.

\end{abstract}
\maketitle                   





\section{Introduction}

Solid-state quantum information systems based on electronic spin or charge 
degrees of freedom offer a number of intrinsic advantages and 
drawbacks. Among the former are the fast operation times, the possibility 
to scale the systems to large size, and the relative ease to integrate 
them into electronic control and read-out circuits. Probably the most serious 
disadvantage is the fact that, in general, solid state devices suffer 
strongly from noise due to internal and external degrees of freedom as well 
as material-specific fluctuations. They lead to relaxation and decoherence processes. 
Hence, one of the major tasks in the field is the understanding and 
control of  noise and decoherence. In this article we will concentrate on 
the analysis of fluctuations which arise due to the discrete nature of the 
electron charge. They lead to what is denoted as {\it shot noise}. Its analysis 
reveals information about electron correlations and entanglement~\cite{QNoise}.

In many circumstances the fluctuations are Gaussian distributed and fully characterized 
by their power spectrum. In order to get further information in the general case, 
e.g. about correlations
and entanglement, the theory of {\it full counting statistics} (FCS)~\cite{Levitov} of
electrons has been developed. It concentrates on the probability distribution for the 
number of electrons transferred through the 
conductor during a given period of time. It yields not only the variance, but all higher 
moments of the charge transfer as well, and thus also contains  
information about rare large fluctuations.  

The FCS has its historical roots in quantum optics, where the counting statistics of 
photons has been used to characterize the coherence of photon sources~\cite{mandelwolf}.  
Photons detected by the photo-counter are correlated in time, reflecting the 
Bose statistics of the particles involved. For electronic 
currents the Fermi statistics is the relevant one, but the
 first attempts to derive the FCS of electrons~\cite{LL} revealed some fundamental 
interpretation problems, related to subtleties of the quantum measurement process. 
We are interested in
the probability that the outcome of a measurement of the charge is $q$. 
According to text-book definitions of projective measurement this quantity 
can be expressed by $\langle n|\delta(q -\hat Q)|n\rangle$, where $\hat Q$ is the 
operator for the transmitted charge, and $|n\rangle $ denotes the quantum state. 
It is tempting to relate the transferred charge to the current operator, 
$\hat Q = \int_0^{t_0}\hat j(\tau)d\tau$. However, in quantum transport problems one has to 
pay attention to the fact that the electric current $\hat j(\tau)$ is an operator, which,
in general, does not commute at different times. This property led to severe 
interpretation problems within the original work. They were resolved in the 
later work of Levitov and Lesovik~\cite{Levitov}, which invokes explicitly 
an extra degree of freedom, namely the detector degree of freedom. 
The paradigm of projective measurement is then applied to this detector degree 
of freedom.

In the meantime the theory of FCS in mesoscopic transport has developed into 
a mature field; some achievement are summarized in Refs.~\cite{QNoise, WBelzig}. 
However, the experimental analysis of the FCS remains a challenge. 
First measurements of the third cumulant of charge transfer through a tunnel 
junction have been reported in Refs.~\cite{Reulet,Bomze} and, very recently, 
the FCS of a semiconductor quantum dot (QD) has been investigated by a real-time 
detection of single-electron tunneling via a quantum point contact~\cite{Ensslin}. 
Furthermore, threshold type of measurements of the FCS using an array of 
over-damped Josephson junctions has been elaborated theoretically in Ref.~\cite{Tobiska}.
 
In this paper we will review our recent results on the FCS of interacting 
electrons in a QD and low-dimensional diffusive conductors.
The QDs are basic constituents of most solid-state quantum information systems.
For example, superconducting single Cooper-pair boxes~\cite{Nakamura} and, 
similarly, a double-dot system formed in a semiconductor
2DEG~\cite{Hayashi} have been  shown to operate as charge qubits.
A metallic QD or single-electron transistor (SET)
can serve as an electro-meter to measure the quantum states
of a charge qubit~\cite{Devoret}.
Since all these devices are based on the charge measurements,
a thorough understanding of the fluctuating properties of charge
becomes crucial for progress in this field. 

Recently further links became apparent between the FCS of electron transport and the 
field of solid-state quantum information processing. 
One of these is related to the use 
of electron entangled states for these purposes. 
Most of the work on entanglement has been performed in optical systems 
with photons~\cite{zeilinger99}, cavity QED systems~\cite{rauschenbeutel00} and 
ion traps~\cite{sackett00}. By now several ideas have been put forward how to generate,
manipulate and detect electronic entangled states~\cite{Beenakker0}. 
It turns out that in solid state systems  entanglement is rather common, 
the nontrivial task remaining its  control and detection. 
For mesoscopic conductors, the prototype scheme of such detection was discussed in 
Ref.~\cite{burkard00}. It has been shown  that the presence of spatially 
separated pairs of entangled electrons, created by some {\em entangler}, can be revealed 
by using a beam splitter and by measuring the correlations of the current fluctuations
in the leads. If the  electrons are injected in an entangled state, 
bunching and anti-bunching of the cross-correlations of current fluctuations
should be found, depending on whether the state is a spin singlet or triplet.
In Ref.~\cite{taddei02} the FCS of entangled electrons has been analyzed in detail.
The FCS depends not only on the scattering properties of the conductor but also on the 
correlations among the electrons that compose the incident beam. 
In Ref.~\cite{Faoro} the Clauser-Horne inequality test for the FCS in the multi-terminal
structures has been proposed in order to detect the entanglement in the source flux of 
electrons. 

A second link is the intrinsic relation between FCS and detector properties of 
a quantum point contact. 
QPCs were suggested as charge detectors in Ref.~\cite{q1} and have been studied 
experimentally in Ref.~\cite{q2}. Recently they have been used as detectors for the 
state of quantum-dot qubits \cite{q3,q32,q34}. 
The operating principle of the QPC detector relies on the dependence of the 
electron current $I$ through the QPC on the state of the two-level system. 
In Ref.~\cite{Averin} the detector properties of the QPC have been calculated beyond  
linear-response for arbitrary energy-dependent transparency and coupling. 
This is the case of interest since for maximum detector sensitivity typical 
measurements are done in the regime of high QPC transparency, $D \simeq 1/2$, and for 
coupling that is not weak \cite{q3}.
It was found that both the back-action dephasing rate $\Gamma$ and the 
measurement rate $W$ are determined by the electron FCS.

A further motivation to study FCS arises from the need to understand the effect of 
interaction on electron transport in disordered low-dimensional conductors.
Disorder enhancement of Coulomb interaction, together with  quantum coherence effects
strongly influence the transport properties of these systems. The FCS
analysis of this long-standing problem provides a deeper insight
into the question. Typically the Coulomb interaction leads to a suppression
of the conductance of mesoscopic samples at low temperatures and bias voltages. 
It has been demonstrated \cite{Zaikin,qdot,array} that 
the strength of this suppression in various types of mesoscopic conductors 
is related to their noise properties. Frequently we find the simple rule: 
the higher the shot noise the stronger is the Coulomb suppression of the conductance. 
The reason is that both shot noise and Coulomb corrections to the transport 
current are manifestations of the discreet nature of the electric charge.
Beyond that, it was shown that the Coulomb correction to the shot noise scales 
with the third moment of the current in the absence of interactions~\cite{GZ1}. 
Furthermore, it was demonstrated by renormalization group studies of the FCS
of short coherent conductors~\cite{Kin_Naz} and of quantum dots~\cite{Bagrets2} 
in the presence of Coulomb interaction that the interaction correction to the 
$n-$th moment of the current is determined by $n+1$-th moment
evaluated in the absence of interaction. Further developing these ideas
we show in the present paper that Coulomb interaction may substantially 
enhance the probability of large current fluctuations in low dimensions,
leading to the appearance of long correlated `trains' in the transferred 
charge. This effect is most pronounced when the system size matches the
dephasing length due to Coulomb interaction. Such coincidence is 
not accidental and comes from the presence of the soft diffusive modes
in the system which strongly renormalize the bare interaction.  

The structure of this article is as follows: In the next section we introduce
some basic definitions and concepts of the FCS in mesoscopic transport. We discuss
the paradigm of quantum measurement by using a spin 1/2 as
galvanometer and consider some simple illustrative examples.
In the main part we concentrate on our own contributions to the field, 
discussing the effects of Coulomb interaction onto the shot noise and FCS 
in interacting quantum dot systems (Section 3) and in 
low-dimensional diffusive interacting conductors (Section 4).

\section{Concepts of FCS}

\begin{figure}[t]
\begin{center}
\includegraphics[width=2.5in]{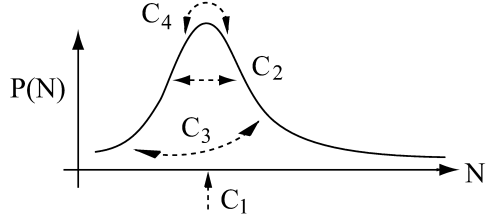}
\end{center}
\caption{The distribution of the number of transmitted electrons $N$.
The mean $C_1$, the variance $C_2$, the skewness $C_3$
and 
the kurtosis $C_4$ characterize 
the peak position, the width, the asymmetry and 
the sharpness of the distribution, respectively.}
\label{fig0}
\end{figure}

We start this section by introducing some definitions and general formulae of 
the FCS approach to mesoscopic transport. The central quantity is the probability 
distribution, $P_{t_0}(N)$, for $N$ electrons to be transferred through the conductor
during a time interval $t_0$. The detection time $t_0$ is assumed to be much
larger than the inverse current frequency $e/I$, which ensures that on average 
$\bar N \gg 1$. This probability distribution $P_{t_0}(N)$ is related to the
cumulant generating function (CGF), ${\cal F}(\chi)$, via a discrete Fourier transform
\begin{equation}
 e^{-{\cal F}(\chi)} = \sum_{N=-\infty}^{+\infty} P_{t_0}(N) e^{i N\chi}.
 \label{E_F}
\end{equation}
The auxiliary variable $\chi$ is usually called \lq\lq counting field".
 From the CGF one readily obtains the \lq\lq cumulants" (irreducible moments)
\begin{equation}
 C_k = \langle\langle N^k\rangle\rangle = -(-i)^k \frac{\partial^k}{\partial\chi^k} {\cal F}(\chi)\bigl|_{\chi=0}.
 \label{IC}
\end{equation} 
The first four of the irreducible moments, defined by
\begin{eqnarray}
 C_1 &=& \bar N = \sum_N  N P_{t_0}(N), \quad C_2 = \overline{(N-\overline{N})^2}, \quad
 C_3 = \overline{(N-\overline{N})^3}, \nonumber \\ 
 C_4 &=& \overline{(N-\overline{N})^4} - 3 C_2^2,
\end{eqnarray} 
denote the mean, variance, asymmetry (\lq\lq skewness") and kurtosis (\lq\lq sharpness"), 
respectively. 
They characterize the peak position, width of the distribution and further details of the 
shape of the distribution $P_{t_0}(N)$, as illustrated in Fig.~\ref{fig0}.

In order to provide a quantum mechanical definition of the CGF 
of electrons we will follow the approach proposed by Levitov and Lesovik~\cite{Levitov}.
The key step is to include the measurement device in the description. As a {\it gedanken} 
scheme a spin-$1/2$ system is used as a galvanometer for the charge detection. 
This spin is placed near the conductor and coupled magnetically to the electric current.
Let the electron system be described by the Hamiltonian ${\cal H}({\bf q, p})$.
We further assume that the spin-$1/2$ generates a vector potential
${\bf a}({\bf r})$ of the form ${\bf a}({\bf r})= \frac{1}{2}\chi\nabla f({\bf r})$.
Here the function $f({\bf r})$ smoothly interpolates between 0 and 1 in the vicinity
of the cross-section at which the current is measured, and $\chi$ is an arbitrary coupling
constant so far. It will be shown below that it plays a role of \lq\lq counting field".
If one further restricts the coupling of the current to the $z$-component of the 
spin then the total Hamiltonian of the system takes the form 
$\hat{\cal H}_\sigma = {\cal H}({\bf q, p - a}\,\hat\sigma_z)$. 

In the semiclassical approximation, when the variation of $\nabla f({\bf r})$ on
the scale of the Fermi wave length $\lambda_F$ is weak, it is possible to linearize
the electron spectrum at energies near the Fermi surface. Thus one arrives at the
Hamiltonian $\hat{\cal H}_\sigma = {\cal H}({\bf q, p}) + \hat{\cal H}_{\rm int}$, where 
\begin{equation}
 \hat{\cal H}_{\rm int} = -\frac{1}{e}\,\hat\sigma_z\int_{-\infty}^{+\infty}d^3{\bf r\,\,\, a(r) j(r)}
 = -\frac{\chi}{2e}\,\hat\sigma_z I_S. \label{Hint}
\end{equation} 
Here ${\bf j(r)}$ is the current density and
$I_S = \int d^3{\bf r\,\,\, j(r)}\, \nabla f({\bf r})$ the total current across 
a surface $S$.
On the quasi-classical level Eq.~(\ref{Hint}) shows that a spin linearly coupled to the 
measured current $I_S(t)$ will precess with a rate proportional to the current.
If the coupling is turned on at time $t=0$ and switched off at $t_0$
the precession angle $\theta = \chi\int_0^{t_0} I_S(t) dt/e$ of the spin
around the $z$-axis is proportional to the transferred charge through the conductor.
In this way the spin-1/2 turns into analog galvanometer.

  To proceed with the fully quantum mechanical description let us consider the evolution
of the spin density matrix $\hat \rho_S(t)$. We assume that initially
the density matrix of the whole system factorizes, 
$\hat\rho = \hat\rho_e\otimes \hat\rho_S(0)$,
with $\hat\rho_e$ being the initial density matrix of electrons. Then the time evolution of 
$\hat\rho(t)$ is given by 
\begin{equation}
 \hat\rho(t) = {\rm Tr}_e \left( e^{-i\hat{\cal H}_\sigma t}\hat\rho\, e^{i\hat{\cal H}_\sigma t}\right),
\end{equation}
where ${\rm Tr}_e$ denotes the trace over electron states. Since, by construction, the 
evolution operator $e^{-i\hat{\cal H}_\sigma t}$ is diagonal in the basis of $\hat\sigma_z$, 
the spin density matrix takes the form 
\begin{equation}
\hat\rho_S(t_0) = \left[
\begin{array}{cc}
\hat\rho_{\uparrow \uparrow }(0) & {\cal Z}(\chi)\hat\rho_{\uparrow \downarrow }(0)  \\
{\cal Z}(-\chi)\hat\rho_{\downarrow \uparrow }(0) & \hat\rho_{\downarrow \downarrow }(0)
\end{array}
\right], \quad 
{\cal Z}(\chi) = {\rm Tr}_e \left( e^{-i {\cal H}_\chi t}\hat\rho_e\, e^{i {\cal H}_{-\chi} t}\right),
 \label{DM}
\end{equation}
where the Hamiltonian 
$\displaystyle {\cal H}_{\chi} = {\cal H}({\bf q, p}) -\frac{\chi}{2e}\, I_S$ 
acts on the electron degrees of freedom only. It becomes clear now that the non-diagonal 
elements of the density matrix~(\ref{DM}) contain the information about the distribution 
of precession angles of the spin during time $t_0$. To make it explicit we use the 
transformation rule of the spin-$1/2$ density matrix corresponding to a rotation around 
the $z$-axis by the angle $\theta$,
\begin{equation}
 {\cal R}_\theta(\hat\rho) = \left[
\begin{array}{cc}
\hat\rho_{\uparrow \uparrow } & e^{i\theta}\hat\rho_{\uparrow \downarrow }  \\
e^{-i\theta}\hat\rho_{\downarrow \uparrow } & \hat\rho_{\downarrow \downarrow }
\end{array}
\right].
\end{equation}
One now can identify ${\cal Z}(\chi)$  with the CGF introduced in Eq.(\ref{E_F}), 
i.e one sets ${\cal Z}(\chi) = e^{-{\cal F}(\chi)}$, and the spin density matrix 
$\hat \rho_S(t_0)$
can be represented as a superposition of the form
\begin{equation}
 \hat \rho_S(t_0) = \sum_{N=-\infty}^{+\infty} P_{t_0}(N) {\cal R}_{\theta = N\chi}(\hat\rho),
\end{equation}
where $P_{t_0}(N)$ has a meaning of the probability to observe the precession at angle 
$\theta = N\chi$. For a classical spin a precession angle $\theta=\chi$
corresponds to a current pulse carrying an elementary electron charge, 
$\displaystyle e = \int_{0}^{t_0} I_S(t) dt$. Using the correspondence principle we 
conclude that the quantity $P_{t_0}(N)$ can be interpreted as the probability of transfer
the multiple charge $N e$. This consideration suggest the use of ${\cal Z}(\chi)$ as
the microscopical quantum mechanical definition for the CGF.
 
\begin{figure}[t]
\begin{center}
\includegraphics[width=6cm]{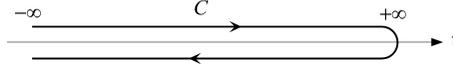}
\end{center}
\caption{Keldysh contour $C$, used for evaluation of the cumulant generating function (CGF) .
} 
\label{fig1}
\end{figure}
 
   Using a cyclic permutation under the trace, ${\rm Tr}_e$, in Eq.~(\ref{DM}) one can
represent ${\cal Z}(\chi)$ in the form of the Keldysh partition function 
\begin{equation}
 {\cal Z}(\chi) = e^{-{\cal F}(\chi)} = \left\langle  T_K 
\exp\left\{ -i\int_C dt\, {\cal H}_\chi(t)\right\}\right\rangle. 
\label{KPF}
\end{equation}
Here the time integration is performed along the Keldysh contour $C$, as shown
in Fig.~\ref{fig1}, and $T_K$ denotes the time ordering operator along the path $C$. 
The average $\langle ... \rangle$ is performed with the non-equilibrium
electron density matrix $\hat\rho_e$.  
The interaction part of the Hamiltonian ${\cal H}_\chi(t)$   reads 
${\cal H}_{\rm int}(t) = \frac{1}{2e} \chi(t) I_S$, where the \lq\lq counting field"
$\chi(t^\pm) = \pm\chi$ is asymmetric on the upper and lower branches of the
Keldysh contour.
   
\begin{figure}[t]
\begin{center}
\includegraphics[width=1.5in]{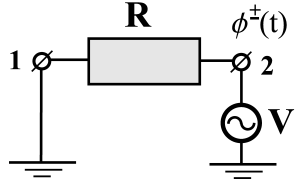}
\end{center}
\caption{ Voltage biased mesoscopic conductor of resistance $R$.
} 
\label{circuit}
\end{figure}

  The definition (\ref{KPF}) for the CGF can be generalized to
obtain the full frequency dependence of the current correlators of arbitrary 
order~\cite{Kin_Naz1}.
Consider a mesoscopic conductor as shown in Fig.~\ref{circuit} coupled to two 
leads such that lead 1 is grounded while  lead 2 is biased with voltage $V(t)$. 
We assume that the current $I(t)$ is measured in the lead 2. This set-up is described
by the interaction Hamiltonian ${\cal H}_{\rm int}(t) = \varphi(t) I(t)/e$
with the phases $\varphi^\pm(t) = \int_{-\infty}^t eV(\tau) d\tau \pm \frac{1}{2}\chi(t)$
defined on the lower/upper branches of the contour $C$, and Eq.~(\ref{KPF}) yields
the generating functional ${\cal Z}[\varphi^+(t), \varphi^-(t)]$ of the 
current fluctuations in the lead 2.
In analogy with the definition~(\ref{IC}) the higher-order derivatives of the 
functional yield the $n$-point irreducible correlation function of currents
\begin{equation}
 e^n C_n(t_1, \dots ,t_n) = 
 (-ie)^n \frac{\delta}{\delta \chi(t_1)} \dots \frac{\delta}{\delta \chi(t_n)} 
\ln{\cal Z}[\varphi^+(t), \varphi^-(t)]\bigl|_{\chi=0}. 
\end{equation}

For illustration we consider an {\bf Ohmic resistor} with resistance $R$ at temperature $T$.
Its CGF is quadratic in $\varphi^\pm(t)$ and reads
\begin{equation}
 {\cal F}_R[\varphi^+(t), \varphi^-(t)] = \frac{1}{2\pi} \frac{R_{\rm Q}}{R}
\int_C dt_1 \int_C dt_2\, \alpha(t_1-t_2) \varphi(t_1) \varphi(t_2), 
\end{equation} 
where $R_{\rm Q} = 2\pi \hbar/e^2$ is the quantum resistance and 
\begin{equation}
 \alpha(t) = \frac{\pi T^2}{2\sinh^2 (\pi t T)}.
\end{equation} 
If the times $t_{1,2}$ lie on  different branches of the contour C the kernel
$\alpha(t)$ is regularized by the shift into the complex plane, $t^\pm \rightarrow t\pm i0$;
otherwise it is understood as principal value.
The quadratic form of ${\cal F}_R$ reflects the Gaussian nature of current fluctuations
in an Ohmic resistor with its well-known properties.

The simplest example of a CGF for non-Gaussian processes is provided 
by a {\bf tunnel junction}. In this case one has~\cite{SZ}
\begin{equation}
 {\cal F}_{\rm T}[\varphi^+(t), \varphi^-(t)] = -4\pi \frac{R_{\rm Q}}{R_{\rm T}}
\int_C dt_1 \int_C dt_2\, \alpha(t_1-t_2) 
\sin^2\left[\frac{\varphi(t_1) - \varphi(t_2)}{2} \right],
\label{CGFtun}
\end{equation}
where $R_{\rm T}$ is a tunnel resistance. 
For a constant applied voltage $eV$ and stationary "counting field" $\chi$
the corresponding CGF reduces to
\begin{equation}
 {\cal F}_{\rm T}(eV,\chi) = -t_0 \left[
\Gamma_+ \left(e^{i\chi}-1\right) + 
\Gamma_- \left(e^{-i\chi}-1\right) \right], \quad \Gamma_\pm = \pm\frac{1}{e^2 R_T}
\frac{eV}{1-e^{\pm eV/T}}.
\label{bipo}
\end{equation}
This result represents the CGF of a bidirectional Poissonian process with rates 
$\Gamma_\pm$
corresponding to uncorrelated tunneling processes of the charge through the junction.
At zero temperature and positive bias voltage the 
second term of Eq.~(\ref{bipo}) disappears. 
Then, performing the inverse Fourier transformation we obtain 
a simple Poisson distribution, corresponding to uncorrelated charge transfer 
\begin{equation}
P(N) = 
\int^{\pi}_{-\pi}
\frac{d \chi}{2 \pi}
\,
\exp ( {\cal F}_T(eV,\chi) -i N \chi )
=
\frac{
\bar{N}^{\, N} {\rm e}^{-\bar{N}}}{N!}, 
\;\;\;\;
\bar{N}=t_0 \Gamma_+.
\end{equation}

\begin{figure}[t]
\begin{center}
\includegraphics[width=1.5in]{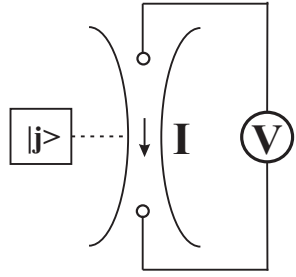}
\end{center}
\caption{ The principal scheme of the qubits readout using the quantum point
contact (QPC). Due to electrostatic coupling the current $I$ driven by the 
external voltage $V$ is controlled by the state $|j\rangle$ of the qubit.
} 
\label{QPC}
\end{figure}

As a further important example we consider a {\bf quantum point contact} (QPC), 
shown in Fig.~\ref{QPC}. It can be used as a quantum detector for the state of a 
quantum dot charge qubit. Its operating principle is based on the property that - 
due to the electrostatic coupling between the dot and the QPC - the scattering matrix 
$
\hat S_j = \left( 
\begin{array}{cc}
 r_j & t_j^* \\
 t_j & r_j^*
\end{array}
\right)
$ 
and thereby the current $I$ of the QPC depend on the 
state $|j\rangle$ of the qubit. 
For a given realization $\hat S$ of the scattering matrix the FCS has been calculated
by Levitov {\it et al.}~\cite{Levitov}, with the result
\begin{equation}
 {\cal F}(\chi) = - t_0 \int \frac{d\epsilon}{2\pi} \ln \det 
\left[ 1 + \hat f \left( \hat S^+(-\chi) \hat S(\chi) - 1\right) \right] 
\label{CGF_Smatrix}
\end{equation} 
where 
$
\hat S(\chi) = \left( 
\begin{array}{cc}
 r_j & t_j^* e^{-i\chi/2}\\
 t_j e^{i\chi/2} & r_j^*
\end{array}
\right),
$  
and $\hat f = {\rm diag} (f_L, f_R)$ is the diagonal density matrix of the leads. 
The explicit evaluation of Eq.~(\ref{CGF_Smatrix}) yields
\begin{equation}
{\cal F}(\chi) = - t_0 \int \frac{d\epsilon}{2\pi} \ln 
\left[ 1 +  f_L(1-f_R) D (e^{i\chi}-1) + f_R(1-f_L) D (e^{-i\chi}-1) \right] \; ,
\end{equation} 
where $D=|t|^2$ is a transmission coefficient. The physical interpretation of this
result is that electrons can be transmitted either forward or backward with 
probabilities $p_{R\leftarrow L} = f_L(1-f_R) D$ and  
$p_{L\leftarrow R} = f_R(1-f_L) D$,
respectively, with the occupation factors accounting for the Pauli principle.

Recently, it has been realized that the quantum detector
properties of the QPC are intrinsically related to its FCS~\cite{Averin}. 
The two basic quantities to be considered are the measurement-induced 
dephasing time of the qubit, and the time needed for the acquisition of information 
about the qubit's state.
Keeping in mind the {\it gedanken} spin-$1/2$ measurement scheme for  charge
detection, described in the beginning of this section, we realize that
in the quantum measurement process by a QPC the role of {\it gedanken} 
galvanometer is played by the qubit (or more generally by a many-level system). 
Then the analog of Eq.~(\ref{DM}) describes the dephasing
rate $\Gamma$ of the qubit's density matrix $\hat\rho(t)$ due to its interaction 
with the electron current in the QPC,
$
 \rho_{jk}(t) = \rho_{jk}(0) \left\langle e^{ i \hat H_k t}  e^{ - i \hat H_j t} \right\rangle
 \simeq \rho_{jk}(0) e^{ - \Gamma_{jk} t} 
$.
Here the average is taken over a stationary state of the QPC and $\hat H_j$ and 
$\hat H_k$ are electron Hamiltonians describing the propagation of electrons with 
scattering matrices $\hat S_j$ and $\hat S_k$. In the long-time 
limit, $t>\hbar/eV$, the decay is exponential with rate
\begin{equation}
 \Gamma_{jk} =  - \int \frac{d\epsilon}{2\pi} \ln \det 
\left[ 1 + \hat f ( \hat S^+_k \hat S_j - 1) \right] \, .
\label{Rate}
\end{equation} 
The analogy with the expression (\ref{CGF_Smatrix}) is striking. 
For a qubit at low temperature, $T\ll eV$, one obtains~\cite{Averin}
\begin{equation}
 \Gamma_{12} = -  \frac{eV}{2\pi\hbar} \ln | t_1 t_2^* + r_1 r_2^* | \, .
\end{equation}

The second aspect of the measurement by the QPC is the rate of information acquisition.
The information about the state $|j\rangle$ of the qubit is
encoded in the probability distribution $P_t^{(j)}(N)$ for $N$ electrons to be
transferred via the QPC, given that the qubit is in the state $|j\rangle$.
The mean value of this distribution, $\bar N_j$, and its width 
$\sqrt{\langle \Delta N_j^2 \rangle}$ grow like $t$ and $t^{1/2}$.
This time dependence implies that only after a certain time, which we denote
as the measurement time $\tau_{\rm meas}$, the two peaks, corresponding 
to different states $|j\rangle$ and $|k\rangle$ emerge from a broadened 
distribution. Quantitatively this time can be defined by
considering the statistical overlap of two distributions,
$M_{jk}(t) = \sum_N \left[ P_t^{(j)}(N) P_k^{(j)}(N)  \right]^{1/2}$.
For long times, $t>\hbar/eV$, the decay should be exponential,  
$M_{jk}(t) \propto \exp\{ - W_{jk} t\}$, with $W_{jk} = 1/\tau_{\rm meas} $
being the measurement rate. As it was shown in Ref.~\cite{Averin} it can
be expressed in terms of the CGF as
\begin{equation}
 W_{jk} = \frac{1}{2t_0}\min{}_{\chi} \left[ {\cal F}_j (\chi) 
+ {\cal F}_k (-\chi) \right]
\end{equation}
In the case of quantum-limited detection the rates $W_{jk}$ and $\Gamma_{jk}$ coincide,
while generally $W_{jk} \leq \Gamma_{jk}$, meaning that dephasing occurs faster than
the information gain. One can show that a QPC can be
operated as quantum limited detector with the rates
\begin{equation}
  \Gamma_{jk} = W_{jk} =  -  \frac{eV}{2\pi\hbar} 
  \ln \left[ (D_j D_k)^{1/2} + (R_j R_k)^{1/2} \right].
\end{equation}
where  $R_{j(k)} = 1 - D_{j(k)}$ are reflection coefficients.

The FCS of tunnel junction and quantum point contact represent the two
simplest generic examples with non-Gaussian current fluctuations.
In both cases the effects of electron-electron interaction were neglected.
This is justified provided the conductance of the system is not too small, 
$G\geq e^2/2\pi\hbar $,  and the length (size) of the system does not exceed 
the inelastic mean free path. Accounting for interaction effects on the FCS
is no trivial task.  This is the case, in particular, when  electrons
have an internal dynamics within the conductor, so that in the derivation of 
${\cal F}(\chi)$ the internal degrees of freedom have to be integrated out. 
Two examples of these calculations are considered in the following sections. 
In Section 3 we discuss the effects of Coulomb interaction onto the shot noise and FCS 
in interacting quantum dot systems,  and in Section 4 we address the
statistic of current fluctuations in the low-dimensional diffusive interacting 
conductors.

 \section{Full Counting Statistics in interacting Quantum Dots}

\begin{figure}[t]
\begin{center}
\includegraphics[width=0.65 \linewidth]{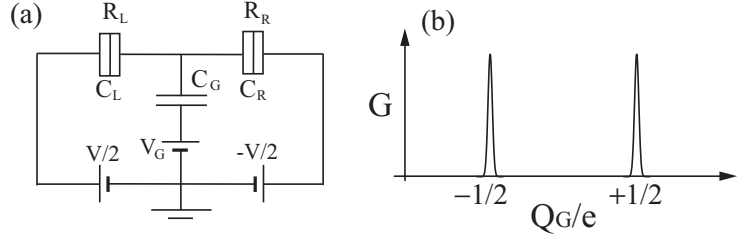}
\end{center}
\caption{
(a) The equivalent circuit of a metallic quantum dot.
(b) The Coulomb oscillations.
}
\label{figure1}
\end{figure}

In this section we review the FCS of a single electron transistor (SET), which is
shown schematically in Fig.~\ref{figure1} (a). It consists of 
a QD with strong local Coulomb interaction, which is coupled via two tunnel junctions 
(left and right) with tunneling resistances $R_{\rm L}$ and $R_{\rm R}$ 
and low capacitances $C_{\rm L}$ and $C_{\rm R}$ to two electrodes (source and drain). 
It is further coupled capacitively via $C_{\rm G}$ to a gate electrode which allows 
- by an applied gate voltage which determines the `gate charge' 
$Q_{\rm G}= C_{\rm G}V_{\rm G}$ - to control the number of electrons inside the QD. 
In the systems of interest the total capacitance 
$C_\Sigma\!=\!C_{\rm L}\!+\!C_{\rm R}\!+\!C_{\rm G}$ is low and, hence, 
the single-electron charging energy $E_C\!=\!e^2/2C_\Sigma$  typically in the 
range of 1K or above. If $E_C$ exceeds the temperature and  applied source-drain bias voltage, $|eV|$,  
the electron transfer through the QD is suppressed (``Coulomb
blockade"). However, the Coulomb barrier can be tuned by the
gate charge. For example, the energy difference between the charge
zero state and the one with one excess electron in the QD depends on $Q_G$ as
$\Delta_{0} = E_C(1 - 2 Q_{\rm G}/e)$. The
voltage drops across the tunnel junctions are $\mu_{\rm r} \!=\!
\kappa_{\rm r} eV$, where $\kappa_{\rm L/R} \!=\! \pm C_{\rm R/L}(C_{\rm
L} \!+\! C_{\rm R})^{-1}$ and the subscript r stands for
either L or R junction. If these voltages
satisfy the condition $\mu_{\rm R} \!<\! \Delta_0 \!<\! \mu_{\rm
L}$ electrons can tunnel sequentially through the island. 
This mechanism leads to the typical oscillating behavior of the conductance as 
a function of the gate charge illustrated in Fig.~\ref{figure1} (b).\\

%
%
The first steps toward a theory of FCS in quantum dots with account
of Coulomb interaction has been performed in Ref.~\cite{AM} where the FCS of 
charge pumping  in the limit of high transmission of the contacts was considered.
Further progress was made by one of the authors and Nazarov~\cite{Bagrets1}, 
who derived the FCS in the frame of a Master equation. 
This approach is valid in the weak tunneling regime, where 
the parameter $g \equiv R_{\rm Q}/R_{\rm T}$, i.e., the ratio between the quantum resistance 
$R_{\rm Q}$ and the effective (parallel) resistance, 
$R_{\rm T}^{-1} = R_{\rm L}^{-1} + R_{\rm R}^{-1}$, is much smaller than unity, $g \ll 1$. 
Applied to a quantum dot in the vicinity of the first conductance peak, 
the CGF is found to differ from a simple Poissonian distribution. Rather it reads
\begin{eqnarray}
{\cal F}^{(1)}(\chi)
  =
t_0 \Gamma  \frac{\sqrt{D(\chi)}-1}{2} \,,
\;\;\;
D(\chi)
 =
1 +
\frac{4\Gamma_{\rm L I} \Gamma_{\rm I R}}{\Gamma^2}(e^{i \chi} \!-\! 1)
+
\frac{4\Gamma_{\rm R I} \Gamma_{\rm I L}}{\Gamma^2}(e^{-i \chi} \!-\! 1) \;.
\label{eqn:CGF-ort}
\end{eqnarray}
Here $\Gamma=\Gamma_{\rm I L}+\Gamma_{\rm I R}+\Gamma_{\rm L
I}+\Gamma_{\rm R I}$, and the rates of electron tunneling 
into/out of the island through the junction r are given by Fermi's golden rule,
\begin{eqnarray}
\Gamma_{  \rm rI/Ir}  = \pm
\frac{1}{e^2   R_{\rm r}} \,
\frac{\Delta_0   -   \mu_{\rm r}}
{e^{\pm(\Delta_0   -   \mu_{\rm r})/T} - 1} \;.
\end{eqnarray}
For special cases the CGF can be simplified: (i) Close to the
threshold of the Coulomb blockade regime, e.g., for $\Delta \approx \mu_{\rm
L}$, and low temperatures $T \ll |eV|$ the tunneling process through
junction L becomes the bottleneck since 
$\Gamma_{  \rm L I} \propto (\mu_{\rm L}-\Delta_0)$ 
is much smaller than $\Gamma_{  \rm IR}$. In this case
the CGF reduces to a Poissonian form,
\begin{equation}
{\cal F}^{(1)}
  \approx
t_0 \, \Gamma_{  \rm L I} \, (e^{i \chi} - 1) \;.
\label{eqn:CGF-threshold}
\end{equation}
(ii)
For a symmetric SET,
$R_{\rm L} \! = \! R_{\rm R}$ and $C_{\rm L} \! = \! C_{\rm R}$,
at the conductance peak, $\Delta_0  = 0$,
one finds for $T=0$ and $eV>0$
\begin{equation}
{\cal F}^{(1)}
  \approx
2 \, \bar{N} \, (e^{i \chi/2} - 1),
\;\;\;\;
e \bar{N}/t_0  =  V/2 (R_{\rm L}   +   R_{\rm R}) \,.
\label{eqn:CGF-deg}
\end{equation}
The extra factor $1/2$ in the exponent leads to a sub-Poissonian value of the Fano factor, 
i.e., ratio between $dc$ power spectrum and  average current
$ S_{II}/2 e \langle I \rangle   \approx   1/2$, indicating that
tunneling processes through the two junctions are correlated.
The distribution function in this case becomes
$P(N) \!=\!
\sum_{N_{\rm L},N_{\rm R}=0}^\infty
P_{\rm P}(N_{\rm L}) \cdot
P_{\rm P}(N_{\rm R})
\, \delta_{N,(N_{\rm L}+N_{\rm R})/2}$,
where the distributions of $N_{\rm L}$ and $N_{\rm R}$ transmitted electrons through 
the L and R junctions have a Poissonian form
$P_{\rm P}(N)
\!=\!
\bar{N}^{\, N} {\rm e}^{-\bar{N}}/N! \,$.
Both are constrained as indicated by the Kronecker $\delta$.

The Master equation approach captures the basic physics of the
strong Coulomb correlations inside the QD, but it neglects non-Markovian
effects, which become important for strongly conducting QDs, i.e., if the dimensionless 
conductance $g$ is no longer small. 
This includes quantum fluctuations of the charge due to co-tunneling, i.e., simultaneous
tunneling of two electrons through two junctions. This process dominates in the
Coulomb blockade regime, i.e. far away from the conductance peaks.
Recently Braggio {\it et al.}~\cite{Braggio} considered these effects in second order 
perturbation theory in $g$ in extension of the theory~\cite{Bagrets1}
using well established
real-time diagrammatic techniques~\cite{SS,Koenig}.

The CGF of a quantum dot in the limit of very strong tunneling, $g \gg 1$, 
has also been considered~\cite{Bagrets2}. 
In this limit the Coulomb blockade almost disappears. Its weak precursor
is caused by quantum fluctuations of the phase, which is the
variable canonically conjugated to the island charge. The
small negative correction to the conductance is logarithmic: $\tilde{g}
\approx g -  2 \ln (\Omega/T),$ where $\Omega \!=\!
1/(R_T C_{\Sigma}) $ is the inverse $RC$
time~\cite{Golubev}.

Several further articles dealt with different setups. In Ref.~\cite{Kindermann1} 
bosonization techniques were used to find the FCS of an open quantum dot coupled to 
reservoirs by single-channel point contacts in the presence of a strong in-plane 
magnetic field. Similarly, the CGF for the generalized two-channel 
Kondo model, which models a QD in the Kondo regime, has been derived~\cite{Komnik}. 
In both cases the authors
succeeded to fully account for Coulomb correlations, but the 
results are limited to a very special, exactly solvable case. 
Despite this work the understanding of the effects of quantum fluctuations 
on the FCS of interacting QDs is far from complete.
In what follows we evaluate CGF for the regime of intermediate strength
conductance.\\

\subsection{FCS OF A SET FOR INTERMEDIATE STRENGTH CONDUCTANCE}

%
%

Here we consider a quantum dot single-electron transistor in the intermediate strength 
tunneling regime, where (introducing for convenience a new dimensionless conductance parameter) 
$\alpha_0 \! \equiv \! g/(2 \pi)^2 < 1$. We assume that the inverse $RC$ time is  
still smaller than the characteristic charging energy,  $\Omega \ll E_C$, which
ensures that the charge-state levels are well resolved. In the vicinity of the conductance peak, precisely
for $|\Delta_0|/E_C\ll 1$,  it is sufficient to restrict the
attention to only two charge states of the quantum dot with
charges differing by $e$. The Hamiltonian can then be mapped onto
the \lq multi-channel anisotropic Kondo model'~\cite{Matveev}.
Introducing a spin-1/2 operator $\hat{\sigma}$, which acts on the
charge states,  we have
\begin{eqnarray}
\hat{H} =
\sum_{\rm r=L,R,I}
\sum_{k n}
\varepsilon_{{\rm r} k} \,
\hat{a}_{{\rm r} k n}^{\dag} \hat{a}_{{\rm r} k n}
+
\frac{\Delta_0}{2} \hat{\sigma}_z
+
\sum_{\rm r=L,R}
\sum_{k k' n}
(
T_{\rm r}
\hat{a}_{{\rm I} k n}^{\dag}
\hat{a}_{{\rm r} k' n}
\hat{\sigma}_{+}
+
{\rm H. c.}).
\label{eqn:H}
\end{eqnarray}
Here $\hat{a}_{{\rm r} k n}^{\dag}$ creates an electron with 
wave vector $k$ and channel index $n$ (including spin) in the left
or right electrode or island (r=L,R,I). The tunneling matrix elements
$T_{\rm r}$ are assumed to be independent of $k$ and $n$.
The junction conductances are
$1/R_{\rm r}
 =  2 \pi e^2 N_{\rm ch} |T_{\rm r}|^2 \rho_{\rm I} \rho_{\rm r}$,
with $N_{\rm ch}$ being the number of channels and $\rho_{\rm r}$ 
the electron DOS. We assume that energy and spin relaxation times
are fast so that electrons are distributed according to 
a Fermi distribution function 
$f(\omega)=1/[\exp(\omega/T)+1]$
both in the island and in the leads.

In the intermediate conductance regime
the main consequence of the quantum fluctuations of the charge
is the renormalization of  system parameters, specifically of
the charging energy and the conductance.
A perturbative two-loop renormalization group analysis for $N_{\rm ch} \gg 1$ predicts
a renormalization of the conductance, $\alpha_0 \to z_0 \alpha_0$, and of the
charging energy, $\Delta_0 \to z_0 \Delta_0$, to depend
logarithmically on the low energy cut-off $\Lambda=\max\{T, \Delta_0\}$~\cite{Matveev}
\begin{eqnarray}
z_0 = \frac{1}{1 + 2 \alpha_0 \ln(E_C/\Lambda)},
\label{eqn:rf}
\end{eqnarray}
Such a conductance renormalization has been confirmed
by experiments~\cite{Joyez}, 
where the observed height of the conductance peaks has been suppressed as $1/\ln T$.

The logarithmic renormalization is typical for Kondo problem, in which one
encounters logarithmic divergences in perturbation theory. 
Likewise, a perturbative treatment of quantum fluctuations in a quantum dot
leads to logarithmic divergences. 
Handling these divergences remains a nontrivial task, 
especially in non-equilibrium transport problem. 
Schoeller and Sch\"on~\cite{SS} have formulated  a real-time diagrammatic approach to this problem.
Summing up a certain class of infinite order diagrams, they
managed to remove the divergences and recover the renormalization factor~(\ref{eqn:rf}). 
Furthermore, they derived non-linear current-voltage characteristics including low-bias Kondo anomalies. 
Recently the second cumulant of the current, i.e. the noise, has been evaluated 
in  lowest~\cite{thielmann1} and second-order perturbation theory~\cite{thielmann2}. 
However, apart from the second-order analysis by Braggio {\it et al.}~\cite{Braggio}, 
the FCS of a quantum dot in the moderate tunneling regime has not been yet analyzed, 
in particular in situations where the finite-order 
perturbation theory fails and infinite order diagrams need to be included. 
Motivated by that, we addressed this problem 
in Ref.~\cite{Utsumi0}, where the FCS of a SET has been evaluated with the use of Majorana  fermion
representation~\cite{Spencer,Coleman,Shnirman}. This formulation 
enabled us to apply Wick's theorem and consequently 
the standard Schwinger-Keldysh approach~\cite{Chou,Kamenev1,Kamenev2}.
 Since practical calculations are rather technical, we will first summarize our main results, 
and postpone the sketch of the derivation to Sec.~\ref{sec:details}.

In the intermediate strength tunneling regime we obtained the CGF in the following form~\cite{Utsumi0}
\begin{eqnarray}
{\cal F}(\chi)
\approx
\frac{t_0}{2 \pi}
\!\! \int \!\! d \omega
\ln [
1
+
T^F\!(\omega)
f_{\rm L}\!(\omega)
\{ 1-f_{\rm R}\!(\omega) \}
({\rm e}^{i \chi}\!-\!1)
+
T^F\!(\omega)
f_{\rm R}\!(\omega)
\{
1-
f_{\rm L}\!(\omega)
\}
({\rm e}^{-i \chi}\!-\!1)
],
\label{eqn:CGF-RTA}
\end{eqnarray}
where
$f_{\rm r}\!(\omega)=f\!(\omega \!-\! \mu_{\rm r})$
is the Fermi distribution function for electrons in the lead 
${\rm r}$. 
This result looks similar to the
Levitov-Lesovik formula for noninteracting systems~\cite{Levitov},
but the effective transmission probability $T^F(\omega)$
accounts for the strong quantum fluctuations of the charge~\cite{Utsumi0,Utsumi},
\begin{eqnarray}
\!\!\!\!\!\!\!\!\!\!\!\!& &
T^F \! (\omega)
= 
(2 \pi)^2
\frac{
\alpha_0^{\rm L}
\alpha_0^{\rm R}
(\omega-\mu_{\rm L})
(\omega-\mu_{\rm R})
}
{|\omega - \Delta_0 - 
\sum_{\rm r=L,R} 
\Sigma^R_{\rm r}(\omega)|^2}
\coth \frac{\omega-\mu_{\rm L}}{2 T} 
\coth \frac{\omega-\mu_{\rm R}}{2 T}
,
\\
\!\!\!\!\!\!\!\!\!\!\!\!& &
\Sigma_{\rm r}^R(\omega)
=
\alpha_0^{\rm r}
\left[
2{\rm Re} \, \psi 
\left( i \, \frac{\omega-\mu_{\rm r}}{2 \pi T} \right)
-
2 \psi \left( \frac{E_C}{2 \pi T} \right)
- \frac{2 \pi T}{E_C}
\right]
- 
i \pi
\alpha_0^{\rm r}
(\omega-\mu_{\rm r})
\coth \frac{\omega-\mu_{\rm r}}{2 T}, 
\label{eqn:se}
\end{eqnarray}
where $\psi$ is the digamma function.
For a symmetric SET at $T\!=\!0$
and  $|\omega| \! \ll \! eV$, the self-energy becomes
$\sum_{\rm r=L,R} 
\Sigma^R_{\rm r}(\omega)
\approx
\alpha_0 \ln ( 2 E_C/eV ) \, \omega - i \Gamma/2$.
 From the real-part of this expression one reproduces 
the logarithmic renormalization~(\ref{eqn:rf}), 
and thus in this sense we accounted for leading logarithms. 
The imaginary part describes the effect of finite life-time of the
charge state of the quantum dot in  non-equilibrium situations.
We also reproduce the average current predicted 
by Schoeller and Sch\"on~\cite{SS}.
Furthermore, the second order expansion in $\alpha_0$,
${\cal F}^{(1)}+{\cal F}^{(2)}$, with
\begin{equation}
{\cal F}^{(2)}(\chi)   =
\partial_{\Delta_0} \{{\rm Re} \Sigma^R_c(\Delta_0)
{\cal F}^{(1)}(\chi) \} + {\cal F}^{\rm cot}(\chi),
\;\;\;
{\cal F}^{\rm cot}(\chi)   =
t_0 \{
\gamma^+ (e^{i \chi} - 1)
+
\gamma^- (e^{-i \chi} - 1) \},
\label{eqn:CGF-cot}
\end{equation}
is consistent with the result of Ref.~\cite{Braggio}.
The first term in the expression for ${\cal F}^{(2)}(\chi)$, 
describes the renormalization (\ref{eqn:rf}) in  
lowest order perturbation theory. 
Corrections of this type for the current were 
 derived earlier in Ref.~\cite{Koenig}. The cotunneling correction to the CGF,
${\cal F}^{\rm cot}$,
describes a bidirectional
Poissonian process governed by the cotunneling rates
\begin{eqnarray}
\gamma^\pm =
2 \pi \,
\alpha_0^{\rm L} \, \alpha_0^{\rm R}
\int  d \omega
\frac{(\omega  -  \mu_{\rm L})
(\omega  -  \mu_{\rm R})}
{(e^{\pm (\omega-\mu_{\rm L})/T}- 1)(1-e^{\mp (\omega-\mu_{\rm R})/T})}
{\rm Re}  \frac{1}{(\omega + i 0 - \Delta_0)^2}.
\nonumber
\end{eqnarray}
This term dominates in the Coulomb blockade regime 
($|\Delta_0| > |eV/2|$ for a symmetric SET) and is consistent with the FCS theory of quasiparticle
tunneling in the presence of many-body interaction~\cite{Levitov2}.

\begin{figure}[t]
\begin{center}
\includegraphics[width=0.75\linewidth]{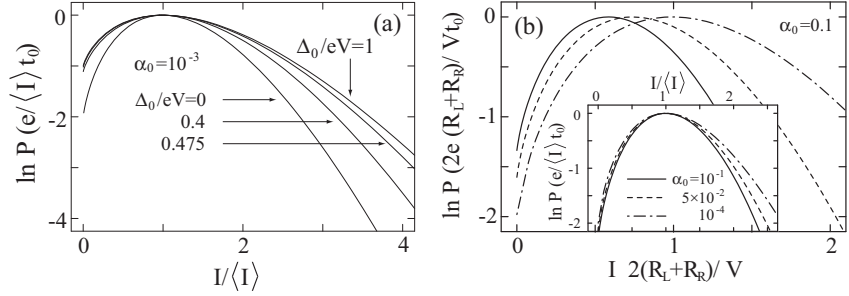}
\end{center}
\caption{
(a) The zero-temperature current distribution
at $eV\!=\!0.2 E_C$
for various values of  $\Delta_0$
(
$R_{\rm L} \!=\! R_{\rm R}$
and
$C_{\rm L} \!=\! C_{\rm R}$).
(b) Plot of $P$ at $\Delta_0  = 0$ for various values
of the conductance versus the current normalized to $V/2 (R_{\rm L}+R_{\rm R})$;
inset: the same distribution normalized to
the average current $\langle I \rangle$.
}
\label{figure2}
\end{figure}

\subsection{NON-MARKOVIAN EFFECTS: RENORMALIZATION AND FINITE \\
LIFETIME BROADENING OF CHARGE STATES}

We present now some results. Figure \ref{figure2}(a) shows the current
($I=eN/t_0$) distribution  for a symmetric SET and for several values of the Coulomb energy barrier $\Delta_0$.
The conductance is chosen to be very small.
As we sweep $\Delta_0$ from the
center of the conductance peak,
$\Delta_0/eV=0$,
to the threshold of the Coulomb blockade regime,
$\Delta_0/eV=0.5$,
the CGF gradually changes from
the correlated Poissonian~(\ref{eqn:CGF-deg})
to the uncorrelated one~(\ref{eqn:CGF-threshold}).
Simultaneously the current distribution widens.
With further increase of $\Delta_0$
one enters into the Coulomb blockade regime, where
the CGF smoothly crosses over to
${\cal F}^{\rm cot}$
and the non-Markovian co-tunneling processes becomes dominant.

As the conductance increases, quantum fluctuations are enhanced.
We find that for $z_0 \Gamma \! \ll \! \Lambda$, [$\Lambda
\!=\!\max(|z_0 \Delta_0|,2 \pi T, |eV|/2)$], the simple expression
for CGF, ${\cal F}^{(1)}$, still holds provided the parameters are
properly renormalized  $\alpha_0\to z_0 \alpha_0$, $\Delta_0\to
z_0 \Delta_0$.
The effect of this renormalization is illustrated in
Fig.~\ref{figure2}(b), where the current distribution for
$\Delta_0  = 0$ is plotted. Since $z_0$ decreases with increasing
$\alpha_0$, the mean value of the current, i.e. the position of the peak in the current distribution,
shifts to lower values. The renormalization effect can be absorbed
if we re-plot the same data with the current (horizontal axis)
normalized by the average current $\langle I \rangle$ rather than
by $V/2(R_{\rm L}+R_{\rm R})$ [inset of Fig.~\ref{figure2}(b)].
However, even after this procedure the three curves do not completely
collapse to a single one. The remaining differences can be
attributed to the non-Markovian effect of the broadening of the
charge states due to their finite life time, which is described by
the imaginary part of the self-energy, Eq.~(\ref{eqn:se}). 
We observe the following trend: 
the probability for current much larger than the average value
is suppressed and 
the current distribution shrinks with increasing $\alpha_0$. 

Let us discuss the lifetime broadening effect quantitatively.
At moderately large voltages, 
$eV   \gg   T_{\rm K}= E_Ce^{-1/2 \alpha_0}/ 2 \pi$, 
and at $T=0$
the real part of the self-energy (\ref{eqn:se}) is negligible and
$\sum_{\rm r=L,R} 
\Sigma^R_{\rm r}(\omega)  \approx - i \pi \alpha_0 eV$.
Then the CGF at $\Delta_0\!=\!0$ reads
\begin{eqnarray}
{\cal F}(\chi) \approx
{\cal F}^{(1)}(\chi)
-
4 \, \bar{q} \, \alpha_0 ({\rm e}^{i \chi}\!-\!1)
+
2 \, \bar{q} \, \pi^2 \alpha_0^2 \,
({\rm e}^{i 3 \chi/2}\!-\!{\rm e}^{i \chi/2})
+O(\alpha_0^3).
\label{eqn:CGFexpansion}
\end{eqnarray}
It is evident from this formula that the higher order cumulants are suppressed
with increasing $\alpha_0$ due to lifetime broadening.

\begin{figure}[t]
\begin{center}
\includegraphics[width=0.6\linewidth]{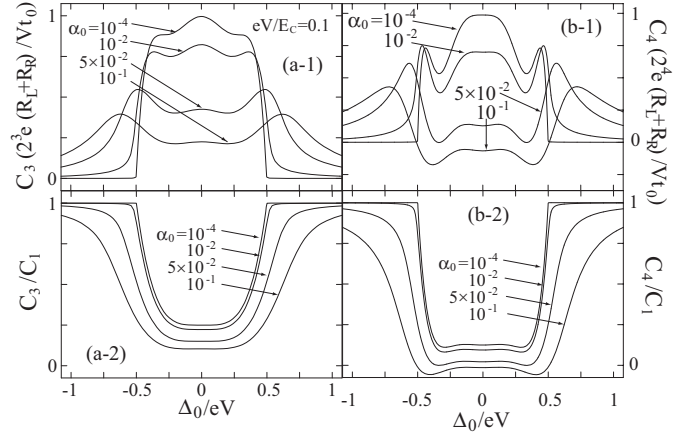}
\end{center}
\caption{
Panels (a-1) and (b-1) are
the skewness and the kurtosis
at $eV\!=\!0.1 E_C$ and 0 K for various conductance.
Panels (a-2) and (b-2) are
those normalized by the average value.
}
\label{figure3}
\end{figure}

Figures~\ref{figure3}(a-1) and (b-1) show the skewness
$C_3$ and the kurtosis
$C_4$ as a function of
$\Delta_0$. A double-peak structure growing with increasing
conductance is observed. In general, we found that higher
cumulants of the current depend on the gate charge in a
complicated way and depend strongly on the conductance. E.g., the
kurtosis even changes its sign for large values of $\alpha_0$.
For the generalized `Fano factors' defined as
$C_3/C_1$ and $C_3/C_1$
[Figs.~\ref{figure2}(a-2) and (b-2)],
we observe a suppression with increasing $\alpha_0$

\subsection{KELDYSH ACTION AND CGF IN MAJORANA REPRESENTATION}
\label{sec:details}

In this section, we sketch our calculations~\cite{Utsumi0,Utsumi}. 
Since within the Schwinger-Keldysh approach the 
calculation of CGF is {\it formally} equivalent to the calculation 
of the partition function~(\ref{KPF}) on the closed time-path $C$ (Fig. \ref{fig1}),  
we can apply the standard field theory methods. 
 From the Hamiltonian~(\ref{eqn:H}) 
we obtain the path-integral representation of the 
Keldysh partition function~(\ref{KPF}) following the standard procedure. 
Tracing out 
the electron degrees of freedom, we obtain  
the effective Keldysh action, which is the sum of two parts: 
the charging part $S_{\rm ch}$ and the
tunneling one $S_t$ --- $S \! \equiv \!S_{\rm ch} \! + \! S_t$.
They read
\begin{eqnarray}
S_{\rm ch}
\!=\!
\int _C \! d t \{ c^* (i \partial_t -\! \Delta_0) c +
\frac{i}{2} \phi \partial_t \phi \},
\;\;\;\;\;
S_t
=- \!\!
\sum_{\rm r=L,R}
\int _C \!\!\! d t d t' \,
c^*(t) \phi(t)
\, \alpha_{\rm r}(t,t') \,
\phi(t') c(t')
+O(T_{\rm r}^4).
\nonumber
\end{eqnarray}
%
Here $c(t)$ and $\phi(t)$ are Grassmann fields which correspond to Dirac fermionic operators $\hat c$
and Majorana fermionic operators $\hat\phi$, respectively. The latter operators are
defined by the following relations:
$\hat{\sigma}_{+}   =   \hat{c}^{\dag} \hat{\phi}$ 
and 
$\hat{\sigma}_{z}   =   2 \hat{c}^{\dag} \hat{c} - 1$.
The action $S_t$ is equivalent to the tunneling 
action~(\ref{CGFtun}), defined earlier.  

A particle-hole Green's function, 
$\alpha_{\rm r}$, describing the tunneling of an electron 
between an electrode r$=$L,R and the island, is expressed in 
the Keldysh space as a $2 \! \times \! 2$ matrix,
\begin{eqnarray}
\tilde{\alpha}_{\rm r}(\omega)
 =
\biggl(
\begin{array}{cc}
0 &   \alpha_{\rm r}^A(\omega) \\
\alpha_{\rm r}^R(\omega) &   \alpha_{\rm r}^K(\omega)
\end{array}
\biggl)
=
-i \pi
\alpha_0^{\rm r}\, \frac{
(\omega  -  \mu_{\rm r})E_C^2}{
(\omega  -  \mu_{\rm r})^2 + E_C^2 }
\biggl(
\!\!
\begin{array}{cc}
0 & -1 \\
1 & \, 2 \coth \frac{\omega-\mu_{\rm r}}{2T}
\end{array}
\!\!
\biggl),
\end{eqnarray}
where, $\alpha_0^{\rm r}\!=\!R_{\rm K}/(4 \pi^2 R_{\rm r})$
is the dimensionless junction conductance.
Here we have introduced a Lorentzian cutoff function to regularize
the UV divergence.
The terms higher than $T_{\rm r}^4$ describe the {\it elastic}
cotunneling process. They can be neglected in the limit of fixed
$\alpha_0$ but  large number of channels, $N_{\rm ch}\gg 1$. In
this case $T_{\rm r}$ scales as $\propto 1/\sqrt{N_{\rm ch}}$ and
thus the terms $\propto T_{\rm r}^4$ give only small corrections 
$\propto 1/N_{\rm ch}$.
In order to derive the CGF, we introduce the counting field $\chi$
by means of the following rotation in the Keldysh space,
\begin{eqnarray}
\tilde{\alpha}^{\chi}_{\rm r}(\omega)
\!=\!
\exp (i \kappa_{\rm r} \chi \mbox{\boldmath$\tau$}_1/2)
\tilde{\alpha}_{\rm r}(\omega)
\exp (-i \kappa_{\rm r} \chi \mbox{\boldmath$\tau$}_1/2).
\label{dph}
\end{eqnarray}
Here $\mbox{\boldmath$\tau$}_1$ is the Pauli matrix.
The CGF takes the following form 
\begin{equation}
{\cal F}(\chi) \! = - \! \ln \! \int {\cal D}
[c^*,c,\phi]
\exp [ i \, S(\chi) ],
\label{CGF}
\end{equation}
where $S(\chi)$ is the effective action
containing the rotated particle-hole Green's function (\ref{dph}).
Tracing out the $c$ fields, we obtain a term of fourth order in $\phi$
in the action, which means that the path integration cannot be performed exactly.
Therefore we proceed in a perturbative expansion in $\alpha_0$ 
and resum a certain class of diagrams. 
Namely we take into account
the contributions from free Majorana Green's function 
and Dirac Green's function with the bubble insertions formed 
by particle-hole Green's (\ref{dph}) function and free Majorana Green's function.

\section{FCS and Coulomb interaction in diffusive conductors}
In this section we consider the FCS of low-dimensional 
diffusive conductors, such as quasi-one-dimensional disordered wires and
two-dimensional disordered films.
It has been appreciated more than two decades ago
that the interplay of interaction and phase coherence effects in
these systems may drastically affect its transport properties~\cite{El_El, Lee}. 
Initially, the conductance was a main object of study, but more recently the
trend moved toward the study of essentially non-equilibrium
phenomena, like the quantum shot noise.

  We consider short diffusive wires and films (with diffusion constant $D$) where the 
Thouless energy $E_{\rm Th}=D/L^2$ is large compared to the applied voltage,
$E_{\rm Th} \gg eV$. In these systems the Fano factor, i.e. the ratio between shot noise 
and current, $S= 2|e|I F$, takes the value $F=1/3$~\cite{Beenakker,Devoret1}. 
The condition for short conductors can be equivalently rewritten
as $\tau_D \ll 1/eV$, where $\tau_D=L^2/D$ is a typical diffusion
time of electron through the system. Such short
conductors are coherent and effectively zero-dimensional so that
all effects of Coulomb interaction come from the external electromagnetic environment. 
It  has been shown recently that the environment modifies the conductance, 
noise~\cite{Zaikin,GZ1} and generally the FCS~\cite{Kin_Naz,Bagrets2}.

Much less is known about the role of 
Coulomb interaction onto the FCS in the quasi- 1D and 2D diffusive systems, when
$\tau_D \gg 1/eV$. Under this condition the inelastic electron-electron scattering
inside the conductor is important. This subject has recently attracted the attention
in Ref.~\cite{Gefen, Pilgram, Mirlin}, where the so-called 
``hot electron'' regime , 
was discussed. It is defined by $\tau_D \gg \tau_E$, with $\tau_E$ being the energy 
relaxation time due to Coulomb interaction, and at the same time $\tau_D \ll \tau_{\rm e-ph}$, 
with $\tau_{\rm e-ph}$ being the electron-phonon relaxation time. These two conditions imply 
that the size of the conductor is larger than the energy relaxation length due to 
electron-electron interaction, but the energy relaxation from the electron
subsystem to phonons is negligible.  In this situation the electron distribution 
function relaxes to the local Fermi distribution with a position dependent electron 
temperature along the conductor. This changes the Fano factor $F$ from $1/3$ to 
$\sqrt{3}/4$~\cite{Hot}, an effect that was confirmed experimentally~\cite{Devoret1}.

\begin{table}
\caption{The electron scattering times for low-dimensional (1D and 2D) diffusive 
conductors, $E=\max\{T,eV\}$.
At $T\leq\tau^{-1}_\phi(V)$, we get $\tau^*=\tau_\phi(V)$.}
\label{tab:table1}
\begin{center}
\begin{tabular}{@{}lccc@{}}
\hline
\hline
d & $1/\tau_E$ & $1/\tau_\phi$ & $1/\tau^*,\,\, T\geq \tau^{-1}_\phi(V)$ \\
\hline
1 & $(E/D)^{1/2}\nu_1^{-1}$ &  $(E^2/D\nu_1^2)^{1/3}$ & $(eV/T)^{1/2}\,\,\tau_E^{-1}(V) $ \\
2 & $E/g$ & $(E/g)\ln g$ & $\ln(eV/T) \,\, \tau_E^{-1}(V)$ \\
\hline
\hline
\end{tabular}
\end{center}
\end{table}

 The microscopic theory~\cite{AA} of electron-electron interaction  in low-dimensional
disordered conductors predicts, however, in addition to $\tau_E$ a further time scale, 
the dephasing time $\tau_\phi$(See Table I). 
Both times are energy dependent and in the limit of good conductors  $g=G/G_Q\gg 1$, which
we wish to consider,  differ parametrically from each other ($\tau_\phi \ll \tau_E$).
It is usually believed~\cite{Aleiner} that classical phenomena described by the 
Boltzmann equation are governed only by the energy relaxation time $\tau_E$,
while the decoherence time $\tau_\phi$ affects essentially quantum-mechanical
phenomena.
Since the FCS is a classical quantity, in the sense that it is proportional
to the number of conducting channels,  one might naively expect that it crosses over 
between the coherent and the ``hot electron" regime
on the scale $\tau_D \sim \tau_E$.

 As we show below the time $\tau_E$ is indeed responsible for a smooth
crossover between the coherent and the ``hot electron" limits if one is interested
in the shot noise and the 3$^d$ cumulant of charge. However, this is not the case for the higher order
cumulants of charge transfer in the shot noise limit $eV\gg T$.  
Moreover, in this limit the smooth crossover in the FCS does not exist.
The Coulomb interaction drastically enhances the probability of 
current fluctuations for short conductors $1/eV \ll \tau_D \ll \tau_E$.
We coined for this range of parameters the term ``incoherent cold electrons''~\cite{DB}.
In what follows we will show that the tail of the current distribution
for such electrons is exponential, $P(I)\sim \exp(-\gamma|I|t_0/e)$. 
The fluctuations are strongest for low temperatures, $T\!\ll\!E_{\rm Th}$, and 
they reach the maximum on the scale $\tau_D \sim \tau_\phi(V)$.
In this case $\gamma\sim g^{-1/2}$ for 1D wire and $\gamma\sim (\ln g/g)^{1/2}$ for 2D film. 
The FCS of this type can be understood as the statistics of a photocurrent which is
generated by electron-hole pairs excited by classical low-frequency fluctuations of 
the electromagnetic field. 
It is remarkable that the time scale of optimal current fluctuations transforms to 
the scale $\tau_\varphi(T)$, known
as a decoherence time in the theory of weak localization~\cite{AA}, 
provided one identifies $eV$ with $T$. 
Therefore, in strongly non-equilibrium
situation the time $\tau_\phi$ rather than $\tau_E$ governs the crossover in the FCS
between the coherent and the ``hot electron" limits.

\subsection{MODEL AND EFFECTIVE ACTION.}
We consider a quasi-one-dimensional (1D) diffusive
wire of length $L$ and a quasi- two-dimensional (2D) film of size $L\times L$, 
with density of states $\nu_d$ per spin, diffusion coefficient $D$ and 
large dimensionless conductance $g = 4\pi \nu_d D L^{2-d} \gg 1$.
They are attached to two reservoirs with negligible external impedance 
which are kept at voltages $\pm V/2$.
The current flows along the $z$ direction and
we concentrate on the incoherent regime, $\max\{eV,T\}\gg E_{\rm Th}$.

  To evaluate the CGF we have used the Keldysh technique and employed
the low-energy field theory of the diffusive transport~\cite{Kamenev} which leads to the action 
\begin{equation}
S[\chi, Q, {\bf A}] = \int d^{\,d}{\bf r}\, {\rm Tr} \left[
\frac{g L^{2-d} }{8} 
\left( {\bf\nabla}Q -i[\hat A , Q] \right)^2 
- 2 \pi\nu_d \Bigl( \partial_t Q\Bigr) \right]  - \frac{i}{8\pi e^2} \int d\,t 
\int \,d^3{\bf r} \Bigr( \dot{\bf A}_1^2  - \dot{\bf A}_2^2 \Bigl) \; .
\label{Action} 
\end{equation} 
Here  $\hat A = {\rm diag}( {\bf A}_1(t, {\bf r}), {\bf A}_2(t, {\bf r}))$ is the 
$2\times 2$ matrix in Keldysh space, where ${\bf A}_{1,2}$ stand for 
fluctuating vector potentials in the conductor. We assume that
${\rm curl}\,{\bf A}=0$, thus neglecting relativistic effects.
The matrix $\hat Q({\bf r}, t_1, t_2)$ accounts for diffusive motion of electrons
and obeys  the semi-classical constraint $\hat Q({\bf r}) \circ \hat Q({\bf r}) = \delta(t_1-t_2)$. 
Boundary conditions are imposed on the field $Q$ in 
the left (L) and right (R) reservoirs~\cite{Belzig},
$Q\bigr|_{{\bf r}=R} = \hat G_R$ and 
$Q\bigr|_{{\bf r}=L} = \hat G_L(\chi) =
e^{i \chi \hat \tau_3/2} \hat G_L e^{-i \chi \hat \tau_3/2}$.
Here $G_{L,R}$ are the Keldysh Green's functions in the leads.

With the action (\ref{Action}) the CGF should be evaluated as a path integral
over all possible realizations ${\bf A}_{1,2}$ and $\hat Q$. 
In general this is a complicated task. However, in the limit 
$1/g\ll 1$  the problem simplifies. We employ the parameterization
$Q = e^{i W} \hat G e^{-i W}$,  $W\hat G + \hat G  W=0$. Here the field $W$ 
accounts for the rapid fluctuations of $Q$ with typical frequencies $\omega\sim eV$ 
and momenta $q\sim\sqrt{eV/D}$, while $\hat G(\epsilon,{\bf r})$ is 
the slow stationary Usadel Green's  function 
varying in space on the scale $\sim L$.
As a first step we integrate out the field $W$ in the Gaussian
approximation to obtain the non-linear action ${\widetilde S}(\chi,\hat G, {\bf A})$ of 
the screened electromagnetic fluctuations  in the media. 
We keep only quadratic terms in ${\widetilde S}$, what is equivalent
to the random phase approximation (RPA). As second step one can integrate out the photon
field $\bf A$ and reduce the problem to an effective action 
$S_{\rm eff}[\chi, \hat G]$.
Then the saddle point approximation,
$\delta S_{\rm eff}[\chi, \hat G]/\delta \hat G = 0$  yields the  kinetic equation
for $\hat G(\epsilon, {\bf r})$.
This program is very similar to that pursued in Ref.~\cite{Kamenev}.

In the universal limit of a short screening radius,
$r^{-1}=(4\pi e^2\nu_3)^{1/2}\gg\sqrt{eV/D}$, 
we get the following result
\begin{equation}
S_{\rm eff}[\chi, \hat G] = \frac{t_0}{8} g L^{2-d} \int d^{\,d}{\bf r}
\int\frac{d\,\epsilon}{2\pi}\, {\rm Tr} 
\left[ {\bf\nabla} \hat G_\epsilon({\bf r}) \right]^2  + \frac{t_0}{2} 
\int \frac{d^{\,d}{\bf r}\, d\omega\,d^{\,d}\bf q}{{(2\pi)}^{d+1}}
{\rm ln}\,\left[
\frac{{\rm Det}||{\cal D}^{-1}_\omega({\bf r, q})||}
{-(D{\bf q}^2)^2-\omega^2} 
\right]
\label{Eff_Action} 
\end{equation} 
where ${\cal D}_\omega$ is a $2\times 2$ matrix operator in Keldysh space 
corresponding to the non-equilibrium diffuson propagator,
\begin{equation}
{\cal D}^{\alpha\,\beta}_\omega({\bf r, q}) 
=\Bigl[ D{\bf q}^2 \,\tau_1^{\alpha\,\beta} + (i/4)\int d\epsilon\, 
{\rm Tr}\left( \gamma^\alpha \gamma^\beta \nonumber 
- \gamma^\alpha \hat G_{\epsilon+\omega/2}({\bf r}) 
\gamma^\beta \hat G_{\epsilon-\omega/2}({\bf r})
\right)
 \Bigr]^{-1}
\label{Diff}
\end{equation}
with $\gamma^0=\hat 1$, $\gamma^1=\hat\tau_3$. 
The first term in 
$S_{\rm eff}$ is due to the diffusive motion of free electrons,
while the second describes the real inelastic electron-electron collisions
with energy transfer $\omega \leq \max\{eV,T\}$.

Minimizing the action $S_{\rm eff}$ under the constraint
$\hat G(\epsilon, {\bf r})^2 = 1$ one obtains a
non-linear matrix kinetic equation for $G(\epsilon,{\bf r})$.
It has a structure of the stationary Usadel equation
\begin{equation}
D\,\nabla\left(\hat G_\epsilon({\bf r})\nabla \hat G_\epsilon({\bf r})\right) = 
\left[\hat{\cal I}_\epsilon({\bf r}), \hat G_\epsilon({\bf r})\right]
\label{KE}
\end{equation}
with the extra matrix collision integral $\hat{\cal I}_\epsilon$ in the r.h.s 
\begin{equation}
\hat{\cal I}_\epsilon({\bf r}) = \frac{i}{8\nu_d}
\sum_{\alpha,\beta} 
\int \frac{d\omega\,d^{\,d}\bf q}{{(2\pi)}^{d+1}}
{\cal D}^{\alpha\,\beta}_\omega({\bf r}, {\bf q}) 
\left[\gamma^\alpha \hat G_{\epsilon-\omega}({\bf r})\gamma^\beta 
+\gamma^\beta \hat G_{\epsilon+\omega}({\bf r})\gamma^\alpha 
\right]
\label{St}
\end{equation}
This kinetic equation 
should be supplemented by the $\chi$-dependent boundary conditions
at the interfaces with the leads, as described after Eq.~(\ref{Action}).
Since we consider the FCS at low frequencies, $\Omega \ll E_{\rm Th}$,
there is no time-dependent term in our kinetic equation
similar to that of the usual time-dependent Usadel equation.
In this limit the collision integral (\ref{St}) guarantees the current conservation,
${\rm div} \,{\bf j}=0$,  where 
${\bf j} \propto \int d\epsilon\, 
{\rm Tr}\left( \hat \tau_3\hat G_\epsilon({\bf r})\nabla \hat G_\epsilon({\bf r})\right)$.
The resulting CGF, ${\cal F}(\chi)$, can be found by evaluating the 
action~(\ref{Eff_Action}) and solving the kinetic equation $G_\epsilon(\chi, {\bf r})$.
In the absence of the field $\chi$ our matrix kinetic equation
reduces to the standard kinetic equation 
with a singular kernel 
$K(\omega) \propto \omega^{\,d/2-2}$ in the collision integral~\cite{AA, Aleiner, Kamenev}.

To derive the action~(\ref{Eff_Action}) we have used a local approximation, i.e. we
neglected gradient corrections proportional to
$(\nabla\hat G\sim 1/L) \ll \nabla W$. 
In this way we incorporate only classical effects of interaction into the FCS. 
The  gradient terms would be responsible for  quantum
corrections to the CGF, coming from frequencies $\omega\gg \max\{T,eV\}$. 
They are small in the parameter $1/g$ and are beyond the scope of this article.

  So far our consideration was rather general. In the following we restrict 
the analysis to the most interesting shot-noise limit, $eV\gg T$. Then the further particular solution
of kinetic equation strongly depends on the relative magnitude of
the diffusion time $\tau_D$ compared to the voltage dependent energy relaxation time $\tau_E(V)$.
In the range of sufficiently high voltages, so that $\tau_D \gg \tau_E(V)$, 
the system is driven into  the ``hot electron'' regime. In this limit the electron
distribution function of electrons has the from of a local
Fermi distribution with position dependent temperature $T({\bf r})$ set by the 
applied voltage $eV$ and differing from the temperature $T$ in the leads.
For smaller voltages, so that  
$1/eV \ll \tau_D \ll \tau_E(V)$, the electrons are described by a 
strongly non-equilibrium two step distribution function, which results from
the weighted average of the Fermi distribution functions in the left
and right leads. We thus call this situation the regime of ``cold electrons''.  
The behavior of the FCS is essentially different in these two regimes and we 
consider them separately in the following subsections.

\subsection{ ``COLD ELECTRON'' REGIME.}
The cold electron regime is defined by relation $E_{\rm Th} \gg 1/\tau_E$.
Under this condition the collision term in the kinetic equation is small, and 
one can obtain the Green's function perturbatively around the coherent solution 
 obeying the Usadel equation 
$\nabla_z\left(\hat G^{\,0}_\epsilon(z)\nabla_z \hat G^{\,0}_\epsilon(z)\right)=0$.
Here $0<z<1$ is a dimensionless coordinate along the current direction. 
This solution was found in Ref.~\cite{Nazarov} and at $T\ll eV$
can be written as
\begin{eqnarray}
&&\hat G^{\,0}_\epsilon(\chi, z) = L_\chi(z)\hat G_L(\epsilon,\chi)+R_\chi(z)\hat G_R(\epsilon)
\label{G0} \\
&&L_\chi(z)=\frac{\sinh(1-z)\,\theta_\chi}{\sinh \theta_\chi },\quad 
R_\chi(z)=\frac{\sinh z\,\theta_\chi }{\sinh \theta_\chi }, \quad
\theta_\chi=\ln(u+\sqrt{u^2-1}), \nonumber
\end{eqnarray}
where $u=2e^{i\chi}-1$ for energies $0<\epsilon<eV$ and $u=1$ otherwise.
In  first order in $\tau_D/\tau_E$ the CGF 
can be found  by substituting $\hat G^0$ in the
action~(\ref{Eff_Action}). The main contribution 
comes from frequencies $T<\omega<eV$.
After some algebra we obtain
\begin{equation}
{\cal F}(\chi)= -(t_0\, g/8\pi)\, \int d\epsilon \,
\theta^2_\chi(\epsilon) + {\cal F}_{\rm Coll}(\chi).
\label{S_chi}
\end{equation}
Here the first term is the CGF of non-interacting electrons, 
and $F_{\rm Coll}(\chi)$ is the correction due to electron-electron
interaction. It reads
\begin{eqnarray}
{\cal F}_{\rm Coll} &=& \frac{t_0 L^{d}}{2}\!
\int\limits_0^1 \, dz\int\limits_{\omega^*}^{eV}
\frac{ d\omega\,d^{\,d} {\bf q} }{ (2\pi)^{d+1} }
{\rm ln}\,\,\left\{ 
1 - \frac{N_\omega\, \omega^2 \Pi(\chi,z)}{(D{\bf q}^2)^2+\omega^2} 
\right\} \label{F_Coll} \\
\Pi(\chi, z) &=&  - 4 \, L_\chi(z) R_\chi(z) e^{i\chi}
\left\{ 1\!-\!L_\chi(z)\!\!-\!\!R_\chi(z) 
- \left[ z L_\chi(z) \!+\!(1-z)R_\chi(z)\right] (e^{i\chi}\!-\!1)\right\}
\nonumber 
\end{eqnarray}
where $N_\omega = (eV/|\omega|-1)$, and
$\omega^* = \max\{E_{\rm Th}, T\}$. The  Thouless energy appears in 
the low frequency cut-off $\omega^*$ due to finite-size effects
while $T$ takes into account the smearing
of a step in the Fermi distribution.
We also note two important properties of the function 
$\Pi(\chi,z)$, namely
(i) $\Pi(i\gamma,z)>0$ for imaginary $\chi=i\gamma$ and
(ii) $\Pi(\chi,z) = -P_2(z)\chi^2 + O(\chi^2)$ at $\chi\ll 1$ where 
$P_2(z) = \frac{\displaystyle 8}{\displaystyle 3} z^2(1-z^2)(1-z+z^2) > 0$.

To estimate the range of validity of the result~(\ref{F_Coll})
we substitute a zero order distribution function 
$f_0(\epsilon) = (1-z) f_F(\epsilon-eV/2) + z f_F(\epsilon+eV/2)$ 
into the collision integral, $f_F(\epsilon)$ being equilibrium Fermi distribution.
Then one estimates the 1$^{\rm st}$ order correction to be
\begin{equation}
\delta f_{(1)}(\epsilon_\pm) \sim  \frac{L^2/D}{\tau_E(V)} \int_{\epsilon_\pm}^{eV} \frac{d\omega}{\omega} 
\left(\frac{eV}{\omega} \right)^{(2-d)/2}
\label{F1}
\end{equation}
if $\epsilon_\pm=|\epsilon\pm eV/2|\ll eV$
and $\epsilon_\pm>\max\{E_{\rm Th},T\}$. By virtue of Pauli's principle
this correction may not exceed unity, $\delta f_{(1)}\leq 1$,
which is true only for 
$\epsilon_{\pm}\geq \epsilon^*$, where the scale $\epsilon^*$ is given by
\begin{equation}
\epsilon^*(V) \simeq \left\{
\begin{array}{cc}
(eV)^2/ g^2\, E_{\rm Th} \; , &  \,\,\mbox {for} \; D=1 \\
 eV \exp\{ - g\,E_{\rm Th}/eV \} \; , & \,\, \mbox {for} \; D=2.
\end{array} \right.
\label{E_star} 
\end{equation}
This result shows that a simple perturbation theory is valid provided
$\epsilon^*<\max\{E_{\rm Th}, T\}$. Resolving this inequality we
obtain the condition $\tau_D < \tau^*$, where the time scale $\tau^*$ is presented in Table 1.
The 1$^{\rm st}$ order perturbation theory breaks down for higher voltages, when  $\tau_D > \tau^*$.  
In this situation we can still obtain the result up to a factor of order of unity 
from Eq.~(\ref{F_Coll}) if we use as cut-off $\omega^*\simeq\epsilon^*$.

The result (\ref{S_chi} - \ref{F_Coll}) with  cut-off $\omega^*=\max\{E_{\rm Th}, T, \epsilon^*\}$
enables us to evaluate all irreducible
cumulants $C_k = -(-i)^k (\partial^k/\partial \chi^k ){\cal F}(\chi)$ 
of a number of electrons transfered.
There is no correction to the current on the classical level. 
The interaction correction to the noise and the
3$^d$ cumulant is small in the parameter $\tau_D/\tau_E$ 
and dominated by inelastic collisions with the energy transfer $\omega\sim eV$.
On the contrary, the leading contribution to the higher order cumulants 
is due to Coulomb interaction and it is 
dominated by quasi-elastic collisions with low energy transfers 
$\omega^* \leq \omega \ll eV$. Up to a numerical constant the result 
is
\begin{equation}
C_{2k, 2k+1} \propto
\frac{\langle N\rangle}{g} 
\Bigl(\frac{\displaystyle eV}{\displaystyle E_{\rm Th}}\Bigr)^{d/2}
\Bigl(\frac{\displaystyle eV}{\displaystyle \omega^*}\Bigr)^{k-1-d/2}, \,\,\mbox {for} \; k\ge 2
\label{cumulants}
\end{equation}
where $\langle N\rangle \gg 1$ is the average number
of electrons transfered and 
$\,\omega^* = \max\{E_{\rm Th}, T,\epsilon^*\}$.

\begin{figure}[t]
\begin{center}
\includegraphics[width=4.5in]{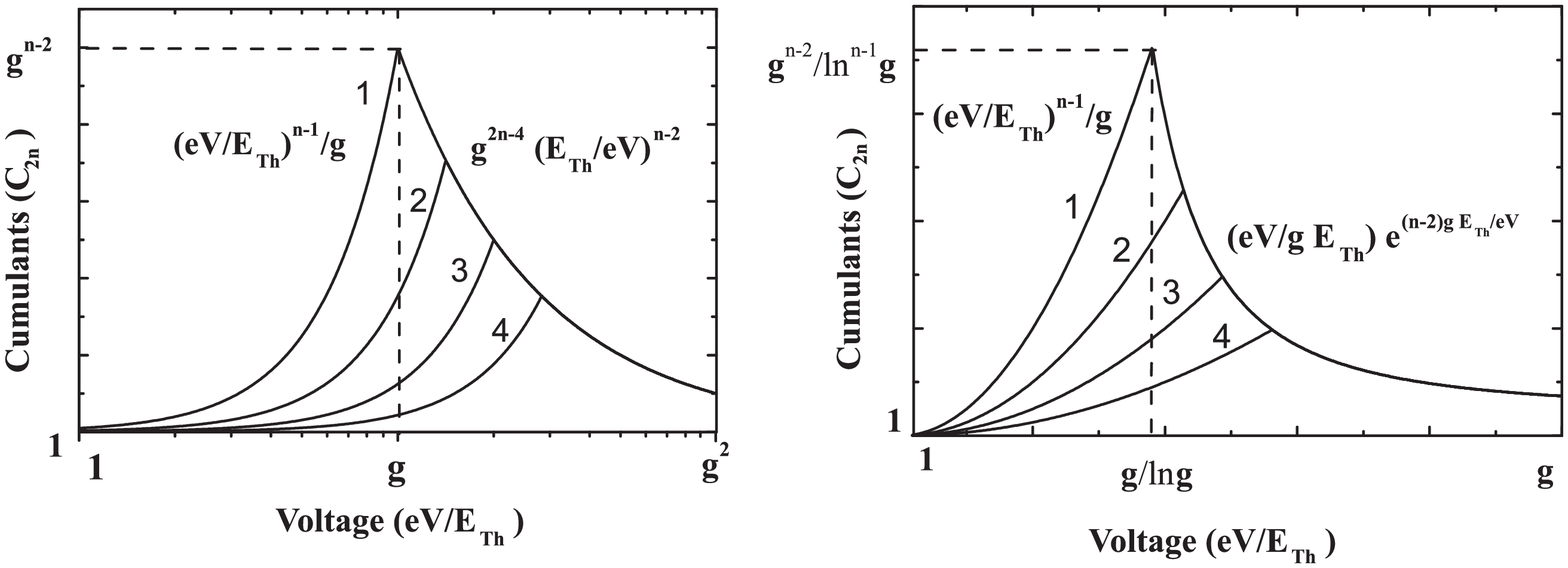}
\end{center}
\caption{ 
Sketch of the voltage dependence of the even cumulants of the charge transfer
$C_{2n}$ for $n>2$
for a 1D diffusive wire (left panel) and a 2D diffusive film (right panel).  
The curves 1,2,3,4 correspond to temperatures $E_{\rm Th}\sim T_1<T_2<T_3<T_4$.
} 
\label{FigCumulants}
\end{figure}

The voltage dependence of the $n^{\rm th}$ cumulant with $n \ge 4$
at different temperatures is sketched in Fig.~\ref{FigCumulants}. 
Eq.~(\ref{cumulants}) shows that the $n+1$st cumulant of the charge transfer 
is parametrically enhanced versus the $k$th one
by the large factor $eV/\omega^*\gg 1$. It also follows from Eq.~(\ref{cumulants}) that
the higher cumulants grow with increasing voltage for
$E_{\rm Th}>1/\tau^*$ and decay for $E_{\rm Th}<1/\tau^*$, where
the new time scale $\tau^*(eV,T)$ is parametrically smaller than $\tau_E$, $\tau^*\ll \tau_E$.
(See Table I).  The current fluctuations are strongest if
$T \leq E_{\rm Th} \sim 1/\tau_\phi(V)$.  
In this case their maximum occurs at $eV/E_{\rm Th}\sim g$ for 1D and
at $eV/E_{\rm Th}\sim g/\ln g$ for 2D.  
To clarify the physical origin of this strong amplification of the current
fluctuations we present below a heuristic interpretation of the result~(\ref{F_Coll})
by relating it to the phenomenon of  photo-assisted shot noise.

 Photo-assisted shot noise has been theoretically predicted by Lesovik and 
Levitov~\cite{Lesovik94}. They considered the mesoscopic scatterer with a single transmission
channel ${\cal T}$ biased by the AC voltage $V(t)=V_\Omega \sin(\Omega t)$ 
(See Fig.~\ref{circuit}). This voltage leads to an oscillating phase 
$\varphi(t) = \int_{-\infty}^{t} eV(\tau) d\tau = \Phi_\Omega \cos\Omega t$
across the conductor with amplitude $\Phi_\Omega = eV_\Omega/\hbar \Omega$. 
It has been shown in Ref.~\cite{Lesovik94} that such a phase modulation results in a 
zero-frequency non-transport shot noise due to the excitation of electron-hole pairs in the leads.
 At low temperatures, $T\ll\hbar\Phi_\Omega$, the noise is
\begin{equation}
 S_2(\omega=0) = \frac{e^2 }{2\pi\hbar} {\cal T}(1-{\cal T}) \sum_{n=-\infty}^{+\infty} 
 | n\Omega|J_n^2(\Phi_\Omega)
\label{PANoise}
\end{equation}
where $P_n = J_n^2(\Phi_\Omega)$ is the probability to excite an electron-hole 
pair with the absorption of $n$ photons, and $J_n(x)$ are Bessel functions.
In the limit of weak phase oscillations, $\Phi_\Omega \ll 1$, the noise is quadratic
in the amplitude, 
$ S_2 = G_Q \hbar\Omega {\cal T}(1-{\cal T}) \Phi_\Omega^2$.
 
The physical origin of the interaction correction to the FCS in diffusive conductors has 
very much in common with the generation of  photo-assisted shot noise. 
Exploiting the path integral formulation of quantum mechanics one can represent 
an interacting electron problem by a picture where a given electron is
moving in a fluctuating electromagnetic field ${\bf A}_{1,2}({\bf r},t)$ created by all other
electrons (with indices $j=1,2$ referring to a forward and 
a backward time evolution operator).  Since the
main effect of the interaction comes from low-frequency fields with $\omega\ll eV$,
the classical part ${\bf A}=({\bf A}_1 + {\bf A}_2)/2$ is of main importance. The field 
${\bf A}_{{\bf q},\omega}$ leads to the excitation 
of electron-hole pairs, which, similar to photo-assisted shot-noise,
produce corrections to the FCS of the form
\begin{equation}
\Delta {\cal F}(\chi, {\bf A}) = -\frac{g L^{2(1-d)}}{4\pi} 
\sum_{\bf q_1, q_2} \frac{ \omega^2 }{(D |{\bf q_1 + q_2}|^2/4)^2+\omega^2} 
\int d^d {\bf r} \, \omega \, \tilde\Pi(\chi, z) {\bf A}_{{\bf q}_1,\omega} 
{\bf A}^*_{{\bf q}_2,\omega} e^{i({\bf q}_1 - {\bf q}_2) {\bf r}}
\label{dF}
\end{equation}
Up to  second order in $\chi^2$ the polarization operator $\tilde\Pi(\chi, z)$
agrees with $\Pi(\chi, z)$ in Eq.~(\ref{F_Coll}). It is important that
this correction is proportional to the total conductance $g$ of the system.
Thus it may become comparable with the non-interacting result for the CGF when the
magnitude of phase fluctuation across the sample becomes of the oder of unity.

In diffusive system with large conductance $g\gg 1$ the fluctuation of ${\bf A}_{{\bf q},\omega}$
are screened and can be considered as Gaussian with an Ohmic spectrum 
${\cal B}(\omega) = \langle |{\bf A}_{{\bf q},\omega}|^2 \rangle = {N_\omega}/({\nu D\omega})$,
where $N_\omega\cong eV/|\omega|$ is a non-equilibrium distribution function 
of electromagnetic  modes.
Thus in order to obtain the interaction contribution to the CGF,
the correction~(\ref{dF}) has to be exponentiated and averaged over these
fluctuations.
Such considerations give exactly the result~(\ref{F_Coll}) with polarization operator 
calculated to accuracy $\chi^2$.

\begin{figure}[t]
\begin{center}
\includegraphics[width=2.5in]{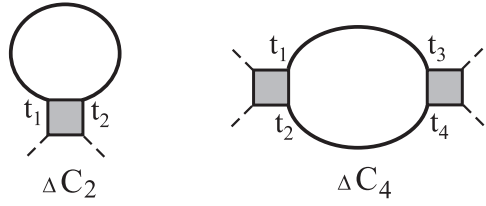}
\end{center}
\caption{ Diagrammatic representation of interaction corrections to the 
shot noise and $C_4$. Shaded blocks correspond to the imaginary part of diffuson,
denoted as $\widetilde{\cal D}_q(t)$, and thick solid lines correspond to the screened
propagator of the electromagnetic fluctuations ${\cal B}(t)$.
Each vertex $t_k$ brings time derivative $\partial_{t_k}$.
} 
\label{Loops}
\end{figure}
  
  It is also instructive to consider the physical picture of photo-assisted
current fluctuations in the time domain. To illustrate the main idea we 
compare the interaction corrections to the 
shot noise $C_2$ and the 4$^{\rm th}$ cumulant $C_4$ for a 1D wire. 
In the time representation they read
\begin{eqnarray}
 \Delta C_2 &=& \int_0^1 P_2(z) \,dz \sum_q 
 \int\widetilde{\cal D}'_{t_1}(q,t_1-t_2){\cal B}'_{t_2}(t_2-t_1) dt_1 dt_2 \\
 \Delta C_4 &=& 6\int_0^1 P^2_2(z) \, dz \sum_q \int \widetilde{\cal D}'_{t_1}(q,t_1-t_2) 
 \widetilde{\cal D}'_{t_4}(q,t_4-t_3)
  {\cal B}'_{t_3}(t_3-t_1)  {\cal B}'_{t_2}(t_2-t_4) dt_1 \dots dt_4 \nonumber
\end{eqnarray}
where
\begin{equation}
 {\cal B}(t) = \int_{1/\epsilon^*}^{+\infty}  \frac{d\omega}{\pi} 
 \frac{N_\omega e^{-i\omega t}}{\nu D \omega},  \quad \mbox{and}\quad 
 \widetilde{\cal D}(q,t) = \int_0^{eV}  \frac{d\omega}{\pi} {\rm Im}
 \left(\frac{1}{ Dq^2 - i\omega}\right)e^{-i\omega t}
\end{equation}
The corresponding Feynman  diagrams are shown in Fig.~\ref{Loops}. The 
correlation time of the diffuson propagator $\widetilde{\cal D}_q(t)$  
is short, given by $\tau_V \sim 1/eV$,
while the photon propagator ${\cal B}(t)$ is strongly non-local in time with  long correlation time 
$\tau\sim 1/{\epsilon^*} \gg \tau_V$. Therefore the time integral in
$\Delta C_2$ is dominated by the short range $|t_1-t_2|\sim \tau_V$ only, and 
the correction to shot noise is small. 
In contrast, when evaluating the correction to $\Delta C_4$ both short, 
$|t_1-t_2|\sim |t_3-t_4| \sim \tau_V$, and long time intervals, 
$|t_1-t_3| \sim |t_2-t_4| \sim 1/\epsilon^*$, are essential. The same structure
holds for higher cumulants $C_{2n>4}$ which are expressed by  one-loop diagrams with
$n$ diffusons and $n$ propagators of the electromagnetic field.
We thus see that a conversion of the long-time electromagnetic field correlations into
the current of electron hole pairs is the reason for an enhancement 
of higher order cumulants in diffusive wires and films. 

  With the results~(\ref{S_chi}, \ref{F_Coll}) we can also explore the current
probability distribution
\begin{equation}
P(I)=\int_{-\pi}^{\pi}\frac{d\chi}{2\pi}\exp\{-\Omega(\chi)\}, 
\quad \Omega(\chi)= -{\cal F}(\chi)+ i (I t_0/e) \chi.
\end{equation}
In the long-time limit, $I t_0/e \gg 1$, this integral can be evaluated within the
stationary phase approximation. For this analysis it is important that the
action ${\cal F}(\chi)$ has two branch points, 
$\chi=\pm i\gamma$, where $\gamma \sim (\omega^*/eV)^{1/2}\ll 1$.
The points $\pm i \gamma$ give two threshold currents, 
$I^{\pm}=(e/t_0) \partial S/\partial\chi\bigl |_{\chi= \pm i\gamma}$, which
read $(I^{\pm} - \langle I\rangle)/\langle I\rangle = \pm \gamma /3$.

\begin{figure}[t]
\begin{center}
\includegraphics[width=2.5in]{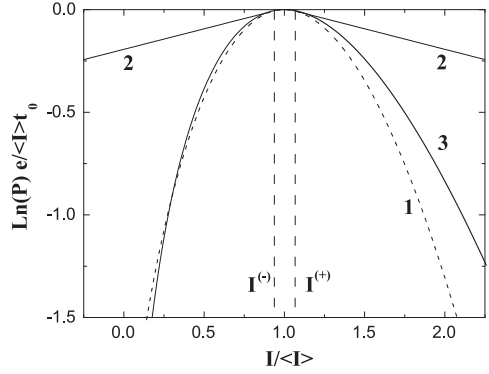}
\end{center}
\caption{ The log of probability to measure the big non-equilibrium current fluctuations
($eV\gg T$). Curve (1), coherent regime;
curve (2), incoherent ``cold electron" regime, $\gamma = 0.2$;
curve (3),  ``hot electron" regime.
} 
\label{FigLnP}
\end{figure}

Provided the fluctuations are small, $I^{-}< I < I^{+}$, the saddle point $\chi^*$ of the function 
$\Omega(\chi)$ lies on the imaginary axis and satisfies the condition 
$|\chi^*|<\gamma$.  Thus with exponentially accuracy the probability distribution becomes 
$P(I)\sim \exp\{-\Omega(\chi^*)\}$.
Due to the smallness of the parameter $\tau_D/\tau_E$ we found
that $P(I)$ deviates only slightly from  the probability $P_0(I)$ of current fluctuations 
in the non-interacting limit.
For larger current fluctuations, $I<I^{-}$ or $I>I^{+}$,
the potential $\Omega(\chi)$
does not have a saddle point any more, and one should use
the contour $C_0$ of a zero phase, ${\rm Im}\,\Omega(\chi)\bigl|_{\chi \in C_0}=0$
for the asymptotic analysis of the integral $P(I)$. This contour 
is pinned by the branch point $\chi=\pm i\gamma$, which yields 
 {\it exponential} tails in the current probability distribution
\begin{equation}
P(I)\approx\exp\{-{\cal F}(\pm\gamma) \mp \gamma I t_0/e \} \; .
\label{Tails}
\end{equation}
The results for the probability distribution are displayed in Fig.~\ref{FigLnP}. 
The Coulomb interaction does not change the
Gaussian fluctuations, but strongly affects the tails of $P(I)$.
They describe long correlated ``trains" in the transfered charge, 
in agreement with our previous discussion on the enhancement
of  higher order cumulants $S_{n \ge 2(D+1)}$.

 One can also relate the statistics (\ref{F_Coll})
to the photocounting statistics studied by Kindermann {\it et al.}~\cite{Kindermann}. 
In that work the FCS of incoherent radiation, which is passed through a highly
transmitting barrier with transmission coefficient ${\cal T}\leq 1$, was studied.
It was shown that a highly degenerate (or classical) source of radiation with
bosonic occupation number $f_\omega \gg 1$ produces  long exponential
tails in the photocounting distribution, $P(n)\propto \exp(-n/f_\omega)$.
The tails of the distribution (\ref{Tails}) are of the same bosonic nature. 
The classical electromagnetic field $A_{\,{\bf q},\omega}$ with
$\omega \ll eV$ and large occupation number $N_\omega\cong eV/|\omega|$
can excite electron-hole pairs with probability 
${\cal P} = \omega^2/((D{\bf q}^2)^2+\omega^2)$. This 
probability plays the role similar to the transmission coefficient ${\cal T}$. It is enhanced
due to the diffusive motion of electrons and can be of order of unity.
The polarization operator $\Pi(\chi, z)$ describes the efficiency of the
conversion of electromagnetic radiation into a current of electron-hole pairs.

\subsection{``HOT ELECTRON" REGIME.}
In the ``hot electron" regime, $E_{\rm Th} \ll 1/\tau_E$, describing the regime of 
high applied voltages or long enough samples, the collision term in the 
kinetic equation~(\ref{KE}) dominates. Thus the saddle point solution of the 
action~(\ref{Eff_Action}) should make the collision integral vanish. 
To find this solution we note the that collision term  
in the action is invariant under the gauge transformation 
$\widetilde{G}_\epsilon({\bf r})= e^{-\hat{K}_{\epsilon}({\bf r})} G_\epsilon({\bf r}) 
e^{ \hat{K}_{\epsilon}({\bf r}) }$. Here
$\hat{K}_{\epsilon}({\bf r}) = \frac{1}{2}\hat\tau_3[\gamma({\bf r}) + 
\beta({\bf r})(\epsilon - \phi({\bf r} ))]$ and 
$\gamma, \beta$ and $\phi$ are arbitrary functions in space. In particular, this
leads to the conservation of the current density,
${\rm div} \,{\bf j}=0$, 
and of the energy flow, ${\rm div}\, {\mathbf j_E}=0$,
where 
${\bf j}_E \propto (2\pi)^{-1}\int\epsilon\, d\epsilon\, 
{\rm Tr}\left( \hat \tau_3\hat G_\epsilon({\bf r})\nabla 
\hat G_\epsilon({\bf r})\right)$.
It is well known that the physical Green function $G(\epsilon,{\bf r})$ with 
a local Fermi distribution 
$f_\epsilon({\bf r}) = [e^{(\epsilon - \phi({\bf r}))/T({\bf r})}+1]^{-1}$ makes the
collision term in conventional kinetic equations vanish. 
Its gauge transform, $\widetilde{G}_\epsilon({\bf r})$, does the same for the
generalized kinetic equation~(\ref{KE}). 

  The four unknown functions
$\phi, \gamma, T$ and $\beta$ can be found from the extremum of the simplified action
$S_{\rm hot}$ which is obtained by substituting $\widetilde{G}_\epsilon({\bf r})$
into the diffusive part of the action $S_{\rm eff}$. For the rest of this discussion 
we restrict ourselves to a 1D wire, since for the  2D geometry shown in Fig.1, all 
results are identical to 1D. We write the action in the form 
$S_{\rm hot}= (2\pi)^{-1}g_0 t \int_0^1 dz S_{\rm hot}(z)$, where
the spatial density $S_{\rm hot}(z)$ reads
\begin{equation}
S_{\rm hot}(z) =  -T(\nabla\gamma-\beta\nabla\phi)^2  + 
 (\nabla\gamma-\beta\nabla\phi)\nabla\phi 
- \frac{\pi^2}{3}T^3(\nabla\beta)^2 + \frac{\pi^2}{6}(\nabla T^2)\nabla\beta 
\label{F_hot}
\end{equation}
Here $T(z)$ and $\phi(z)$ have a meaning of a local temperature and  
chemical potential, while $\beta(z)$ and $\gamma(z)$ are their quantum conjugate counterparts.
The action~(\ref{F_hot}) has to be minimized subject to the boundary conditions
$\phi(z)\bigl|_{z=0,1}=\pm eV/2$, $T(z)\bigl|_{z=0,1}=T$, 
$i\gamma(0)=\chi$ and $\gamma(1)=\beta(0)=\beta(1)=0$. 

The action
$S_{\rm hot}$ possesses 4 integrals of motion. They 
are the physical current $J=\partial S_{\rm hot}/\partial\nabla\gamma$,
the ``quantum" current $M=\partial S_{\rm hot}/\partial\nabla\phi$, 
the energy current 
$J_E = J\phi - \frac{2\pi^2}{3}T^3\nabla\beta + \frac{\pi^2}{6}(\nabla T^2)$, 
and the spatial density of the action $S_{\rm hot}(z)$. Performing
the Legendre transform ,
we can reduce the task to the boundary value problem for two functions $T(z)$ and $\beta(z)$.
Since it appears not to be possible to obtain an analytic solution of these equations f
or non-vanishing $\chi$, we solved them numerically. The results for the probability
distribution $P(I)$ are shown in Fig.~\ref{FigLnP}. As in the previous section it
can be evaluated using the saddle point approximation. We can see from Fig.~\ref{FigLnP} 
that the probability
of positive current fluctuations, $\Delta I>0$, is enhanced in the ``hot electron"
limit as compared to the coherent regime, while the probability of 
negative fluctuations, $\Delta I<0$, is affected to a lesser extent.
The action~(\ref{F_hot}) is equivalent to the actions
of Refs.~\cite{Pilgram, Mirlin} under the appropriate change of variables. 
A further increase of the voltage or the sample size will eventually bring the
system into the macroscopic regime, $L\gg L_{\rm e-ph}$. 
The conductor in this case displays 
only Nyquist noise $S = 4 k_{\rm B}T/R$, while higher order cumulants vanish 
and the probability of current fluctuations becomes Gaussian~\cite{Mirlin}. 

\section{Summary}

To summarize, we have studied the full current statistics (FCS) of charge transfer in 
two important examples of the mesoscopic conductors taking into account the effects of 
Coulomb interaction.  First,  we derived the FCS for a single-electron transistor 
with Coulomb blockade effects in the vicinity of a conductance peak.  
Quantum fluctuations of the charge are taken into account 
by a summation of a certain subclass of diagrams,
which corresponds to the leading logarithmic approximation. 
In lowest order in the tunneling strength our results reproduce 
the `orthodox' theory, while in second
order they account for renormalization and cotunneling effects. 
We have shown that in non-equilibrium situations quantum
fluctuations of the charge induce lifetime broadening for the
charge states of the central island.  
An important consequence is the reduction of the probability 
for currents much larger than the average value.

We further investigated the effect of Coulomb interaction onto the FCS in 
one- and  two-dimensional diffusive conductors. We have found that Coulomb
interaction essentially enhances the probability of rare current fluctuations
for short conductors, $1/eV \ll \tau_D \ll \tau_E$, 
with $\tau_D$  and $\tau_E$ being the diffusion and energy relaxation times.  
The fluctuations are strongest at low temperatures, $T\ll 1/\tau_D$, and they reach
the maximum when the sample size matches the voltage dependent dephasing length
due to Coulomb interaction.
We have shown that tails of the probability distribution of the transfered charge are 
exponential and they arise from the correlated fluctuations of the current 
of electron-hole pairs which are excited by the classical low-frequency 
fluctuations of the electromagnetic field in the media.

\begin{acknowledgement}
This work is part of a research network of the Landesstiftung 
Baden-W\"urttemberg gGmbH.
\end{acknowledgement}

\end{document}